\documentclass[11pt,a4paper]{article}
\pdfoutput=1
\usepackage{jheppub}

\usepackage{natbib}
\usepackage{subfig}

\usepackage{multirow}

\newcommand{\Lumint}{{\cal L}_{\rm int}}

\def\lsim{\mathrel{\rlap{\raise 2.5pt \hbox{$<$}}\lower 2.5pt\hbox{$\sim$}}}
\def\gsim{\mathrel{\rlap{\raise 2.5pt \hbox{$>$}}\lower 2.5pt\hbox{$\sim$}}}

\renewcommand{\Re}{{\rm Re\thinspace}}
\renewcommand{\Im}{{\rm Im\thinspace}}

\allowdisplaybreaks 

\title{Probing the charged Higgs boson at the LHC\\ in the CP-violating type-II 2HDM}
\author[a]{L. Basso,}
\author[b]{A. Lipniacka,}
\author[c,d]{F. Mahmoudi,}
\author[e,f]{S. Moretti,}
\author[b,c]{P. Osland,}
\author[g]{G. M. Pruna,}
\author[b]{M. Purmohammadi}

\affiliation[a]{Albert-Ludwigs-Universit\"at - Fakult\"at f\"ur Mathematik und Physik, D-79104 Freiburg i.\ Br., Germany}
\affiliation[b]{Department of Physics and Technology, University of Bergen,
Postboks 7803, N-5020 Bergen, Norway}
\affiliation[c]{CERN Theory Division, Physics Department, CH-1211 Geneva 23, Switzerland}
\affiliation[d]{Clermont Universit{\'e}, Universit\'e Blaise Pascal, CNRS/IN2P3,\\
LPC, BP 10448, 63000 Clermont-Ferrand, France}
\affiliation[e]{School of Physics \& Astronomy, University of Southampton,\\
Highfield, Southampton SO17 1BJ, UK}
\affiliation[f]{Particle Physics Department, Rutherford Appleton Laboratory, \\Chilton,
Didcot, Oxon OX11 0QX, UK}
\affiliation[g]{TU Dresden, Institut f\"ur Kern- und Teilchenphysik,
Zellescher Weg 19, D-01069 Dresden, Germany}

\emailAdd{lorenzo.basso@physik.uni-freiburg.de}
\emailAdd{Anna.Lipniacka@ift.uib.no}
\emailAdd{mahmoudi@in2p3.fr}
\emailAdd{S.Moretti@soton.ac.uk}
\emailAdd{Per.Osland@ift.uib.no}
\emailAdd{Giovanni\_Marco.Pruna@tu-dresden.de}
\emailAdd{Mahdi.PurMohammadi@ift.uib.no}

\abstract{
We present a phenomenological study of a CP-violating two-Higgs-doublet Model with type-II Yukawa 
couplings at the Large Hadron Collider (LHC). In the light of recent LHC data, we focus on the parameter space 
that survives the current and past experimental constraints as well as theoretical bounds on the model. Once 
the phenomenological scenario is set, we analyse the scope of the LHC in exploring this model through the discovery of a 
charged Higgs boson produced in association with a $W$ boson, with the former decaying into the lightest neutral Higgs 
and a second $W$ state, altogether yielding a $b\bar b W^+W^-$ signature, of which we exploit the $W^+W^-$ semileptonic 
decays.}


\begin{document}

\hfill {\tt CERN-PH-TH/2012-068}

\hfill {\tt FR-PHENO-2012-005}

\hfill {\tt SHEP-12-06}
\vspace*{-1.5cm}

\maketitle

\section{Introduction}
\label{Sec:Intro}
\setcounter{equation}{0}
One of the main tasks of the Large Hadron Collider (LHC) experiments is to study and understand the mechanism of Electroweak Symmetry Breaking (EWSB). 
Recently, the ATLAS and CMS collaborations released the results of the search for the Higgs bosons with more than 10~fb$^{-1}$ data collected at 7~TeV in 2011 and at 8~TeV in 2012 \cite{Aad:2012xx,Chatrchyan:2012xx}. Both experiments have recorded an excess of events above the expected background in different decay channels (mainly $\gamma\gamma$, $ZZ$ and $W^+W^-$) at a mass near 125 GeV. The excess is compatible with the Standard Model (SM) Higgs boson. Complementary evidence is also provided by the updated combination of the Higgs searches performed by the CDF and D0 collaborations at the Tevatron~\cite{Tevatron:2012zzl}.

Investigating the Higgs mechanism in the framework of the SM constitutes a major effort \cite{Djouadi:2005gi}. However, the minimal choice of Higgs sector is (so far) arbitrary. Even if the existence of a scalar resonance, compatible with an SM-like Higgs, has already been uncovered by the current data, one must take advantage of the unique opportunity to test the phenomenology of more complicated Higgs models.

Much effort has been dedicated over the years to the study of extended Higgs sectors. In this paper we consider the two-Higgs-doublet Model (2HDM) with type-II Yukawa sector. This model is one of the most popular extensions of the Higgs sector due to its strong connection with a tree-level Minimal Supersymmetric Standard Model (MSSM) \cite{Gunion:1989we,Djouadi:2005gj}, which is one of the most accredited proposals for solving some theoretical inconsistencies of the SM. 
As is well known, the Higgs sector of the MSSM is quite well constrained in terms of the number of free parameters on which the
masses depend \cite{Arbey:2011ab,Arbey:2011aa,Heinemeyer:2011aa,Hall:2011aa,Draper:2011aa,Carena:2011aa,Cao:2011sn,Cao:2012fz,Christensen:2012ei,Ghosh:2012dh,Brummer:2012ns,Carena:2012gp,CahillRowley:2012rv,Benbrik:2012rm,Arbey:2012dq,Akula:2012kk}. It is possible that the Higgs sector lies in a lower mass range than the superpartners of the SM particles.
In this regard, the 2HDM should be explored as an effective low-energy MSSM-like Higgs sector.

While a tree-level MSSM Higgs sector is strictly CP-conserving (no mixing is allowed between the scalar and pseudo-scalar Higgs components), it has been shown that a CP-violating effective Higgs sector could be produced by loop corrections under specific circumstances \cite{Carena:2000yi}. Accordingly, in this paper we adopt a bottom-up approach by considering a CP-violating 2HDM with type-II Yukawa couplings.

Due to its complicated Yukawa structure, a CP-violating parameter space must be carefully constrained by theoretical arguments and experimental data. Therefore, our first aim is to provide a detailed analysis of the allowed parameter space in the light of recent LHC results. We will show that very little CP-conserving parameter space survives these data.
This exploration of the allowed parameter space has been addressed recently by several authors, from different points of view \cite{Ferreira:2011aa,Burdman:2011ki,Cervero:2012cx,Barroso:2012wz,Arhrib:2012yv}. Where there is overlap, we compare our results with those obtained by these authors.

Regarding phenomenology, the only way to unambiguously probe the existence of a Higgs sector with two doublets arises through the discovery of a charged Higgs boson, since this particle is the hallmark of such a structure of the Higgs sector. Hence, our second aim is to profile a charged Higgs boson in the surviving parameter space via a detailed study of its production cross-section and decay Branching Ratios (BRs).

Then, our third aim is to study the scope of the LHC in discovering a charged Higgs state. In this respect, it is well known that the production of a single charged Higgs state at a hadron collider proceeds in association with either top/bottom quarks or scalar/vector bosons \cite{Moretti:2001pp,Moretti:2002ht}. By taking into account the recent experimental excess observed by ATLAS and CMS \cite{Aad:2012xx,Chatrchyan:2012xx} and the Tevatron \cite{Tevatron:2012zzl} (i.e., corresponding to a light Higgs with a mass of $\approx125$ GeV), we propose a search strategy for a charged Higgs boson produced in association with a $W$ boson and decaying into a $b\bar bW$ final state. In particular, we show that an appropriate choice of the selection cuts would allow the discovery of such a particle despite the considerable $t\bar t$ dominated background.

This paper is organised as follows: in section~\ref{Sec:Model} we give an overview of the considered model, in section~\ref{Sec:bounds} we analyse the allowed parameter space in the light of both theoretical and experimental constraints, in section~\ref{Sec:Results} we present the main phenomenological results, in section~\ref{Sec:Scenarios} we briefly comment on possible future developments, and in section~\ref{Sec:Conclusions} we present our conclusions.  In the appendix we discuss the decoupling limit for the CP-violating type-II 2HDM.

\section{The model}
\label{Sec:Model}
\setcounter{equation}{0}
We describe here our parametrisation of the 2HDM with type-II Yukawa couplings. 
The Higgs sector is defined by the presence of two Higgs doublets, with one ($\Phi_2$) field coupled to the $u$-type quarks,
and the other ($\Phi_1$) to the $d$-type quarks and charged leptons \cite{Gunion:1989we}.

We take the 2HDM potential to be
\begin{align}
\label{Eq:pot_7}
V&=\frac{\lambda_1}{2}(\Phi_1^\dagger\Phi_1)^2
+\frac{\lambda_2}{2}(\Phi_2^\dagger\Phi_2)^2
+\lambda_3(\Phi_1^\dagger\Phi_1) (\Phi_2^\dagger\Phi_2) \nonumber \\
&+\lambda_4(\Phi_1^\dagger\Phi_2) (\Phi_2^\dagger\Phi_1)
+\frac{1}{2}\left[\lambda_5(\Phi_1^\dagger\Phi_2)^2+{\rm h.c.}\right] \\
&-\frac{1}{2}\left\{m_{11}^2(\Phi_1^\dagger\Phi_1)
\!+\!\left[m_{12}^2 (\Phi_1^\dagger\Phi_2)\!+\!{\rm h.c.}\right]
\!+\!m_{22}^2(\Phi_2^\dagger\Phi_2)\right\}. \nonumber
\end{align}

The $Z_2$ symmetry will be respected by the quartic terms (there are
no $\lambda_6$ or $\lambda_7$ terms), and
Flavour-Changing Neutral Currents (FCNCs) are constrained \cite{Glashow:1976nt}.

We parametrise the Higgs fields as
\begin{equation}\label{Eq:Higgs_goldstones}
\Phi_1=
\left(
\begin{array}{c}
- s_\beta H^+ \\
\frac{1}{\sqrt{2}} [v_1 + \eta_1 - i s_\beta \eta_3]
\end{array}
\right), \qquad
\Phi_2 =
\left(
\begin{array}{c}
c_\beta H^+ \\
\frac{1}{\sqrt{2}} [v_2 + \eta_2 + i c_\beta \eta_3 ]
\end{array}
\right).
\end{equation}
The real and
non-negative Vacuum Expectation Values (VEVs) for the Higgs doublets
are $v_1=v c_\beta$ and $v_2= v s_\beta$, 
with $c_\beta=\cos\beta$ and $s_\beta=\sin\beta$, and the ratio defines
\begin{equation}
\tan{\beta}=\frac{v_2}{v_1}.
\end{equation}

CP violation is allowed, and it is realised by means of the fact that $\lambda_5$ and $m_{12}^2$ are complex numbers.
All three neutral states will then mix, with the physical Higgs particles $H_i$ ($i=1,2,3$) related to the weak fields
$\eta_j$ ($j=1,2,3$) of Eq.~(\ref{Eq:Higgs_goldstones}) by
\begin{equation} \label{Eq:R-def}
\begin{pmatrix}
H_1 \\ H_2 \\ H_3
\end{pmatrix}
=R
\begin{pmatrix}
\eta_1 \\ \eta_2 \\ \eta_3
\end{pmatrix}.
\end{equation}
In terms of the non-diagonal mass-squared matrix ${\cal M}^2$, determined from second derivatives of the above
potential, we have
\begin{equation}
\label{Eq:cal-M}
R{\cal M}^2R^{\rm T}={\cal M}^2_{\rm diag}={\rm diag}(M_1^2,M_2^2,M_3^2).
\end{equation}
The $3\times3$ mixing matrix $R$ governing the neutral sector will
be parametrised in terms of the angles $\alpha_1$, $\alpha_2$ and $\alpha_3$
as in \cite{Khater:2003wq,Accomando:2006ga}:
\begin{equation}
R=
\begin{pmatrix}
c_1\,c_2 & s_1\,c_2 & s_2 \\
- (c_1\,s_2\,s_3\!+\!s_1\,c_3) 
& c_1\,c_3\!-\!s_1\,s_2\,s_3 & c_2\,s_3 \\
- c_1\,s_2\,c_3\!+\!s_1\,s_3 
& - (c_1\,s_3\!+\!s_1\,s_2\,c_3) & c_2\,c_3
\end{pmatrix}
\end{equation}
where $c_1=\cos\alpha_1$, $s_1=\sin\alpha_1$, etc., and
\begin{equation}
-\frac{\pi}{2}<\alpha_1\le\frac{\pi}{2},\quad
-\frac{\pi}{2}<\alpha_2\le\frac{\pi}{2},\quad
0\le\alpha_3\le\frac{\pi}{2}.
\end{equation}
For these
angular ranges, we have $c_i\ge0$, $s_3\ge0$, whereas $s_1$ and $s_2$ may be
either positive or negative.  We will use the terminology ``general 2HDM'' as
a reminder that CP violation is allowed.

With all three masses different, there are three limits of {\it no} CP-violation, i.e., with two Higgs bosons that are 
CP-even and one that is odd.
In the above notation, the three limits are \cite{ElKaffas:2007rq}: 
\begin{alignat}{2} \label{Eq:CP-cons}
&\text{$H_1$ odd:} &\quad
&\alpha_2\simeq\pm\pi/2,\ \alpha_1, \alpha_3 \text{ arbitrary}, \nonumber \\
&\text{$H_2$ odd:} &\quad
&\alpha_2=0,\ \alpha_3=\pi/2,\ \alpha_1 \text{ arbitrary}, \nonumber \\
&\text{$H_3$ odd:} &\quad
&\alpha_2=\alpha_3=0,\ \alpha_1 \text{ arbitrary}.
\end{alignat}

In the general CP-violating case, the neutral sector is conveniently
described by these three mixing angles, together with two masses $(M_1, M_2)$,
$\tan\beta$ (the ratio between the two Higgs VEVs) and the parameter $\mu^2=\Re m_{12}^2/(2\cos\beta\sin\beta)$.
From Eq.~(\ref{Eq:cal-M}), it follows that
\begin{equation}
\label{Eq:calM-RMsqR}
({\cal M}^2)_{ij}=\sum_k R_{ki} M_k^2 R_{kj}.
\end{equation}
When CP is violated, both $({\cal M}^2)_{13}$ and $({\cal
M}^2)_{23}$ will be non-zero. In fact, they are related by
\begin{equation} \label{Eq:rel13-23}
({\cal M}^2)_{13}=\tan\beta({\cal M}^2)_{23}.
\end{equation}

From these two equations, (\ref{Eq:calM-RMsqR}) and (\ref{Eq:rel13-23}), we
can determine $M_3$ from $M_1$, $M_2$, the angles
$(\alpha_1,\alpha_2,\alpha_3)$ and $\tan\beta$ \cite{Khater:2003wq}:
\begin{equation} \label{Eq:M_3}
M_3^2=\frac{M_1^2R_{13}(R_{12}\tan\beta\!-\!R_{11})
\!+\!M_2^2R_{23}(R_{22}\tan\beta\!-\!R_{21})}
{R_{33}(R_{31}-R_{32}\tan\beta)}
\end{equation}
where we impose $M_1\le M_2\le M_3$.

Providing also $M_{H^\pm}$ and $\mu^2$, all the $\lambda$'s are consequently determined. 
Since the left-hand side of (\ref{Eq:calM-RMsqR}) can be expressed
in terms of the parameters of the potential (see, for example,
\cite{ElKaffas:2006nt}), we can solve these equations and obtain the
$\lambda$'s in terms of the rotation matrix, the neutral mass eigenvalues,
$\mu^2$ and $M_{H^\pm}$.
The explicit expressions are given in Ref.~\cite{ElKaffas:2007rq}.

The interest in allowing for CP violation lies in the fact that it may
be helpful for baryogenesis \cite{Riotto:1999yt}. Also, from a more pragmatic
point of view, it opens up a bigger parameter space, and allows
certain couplings to be larger.

\subsection{Yukawa and gauge couplings}
For the type-II 2HDM, and for the third generation, the neutral-sector
Yukawa couplings are (assuming all fields incoming):
\begin{alignat}{2}  \label{Eq:H_j_Yuk}
&H_j  b\bar b: &\qquad
&\frac{-ig\,m_b}{2\,m_W}\frac{1}{\cos\beta}\, [R_{j1}-i\gamma_5\sin\beta R_{j3}], 
\nonumber \\
&H_j  t\bar t: &\qquad
&\frac{-ig\,m_t}{2\,m_W}\frac{1}{\sin\beta}\, [R_{j2}-i\gamma_5\cos\beta R_{j3}].
\end{alignat}
Likewise, the charged-Higgs couplings are \cite{Gunion:1989we}
\begin{alignat}{2}  \label{Eq:Yukawa-charged-II}
&H^+ b \bar t: &\qquad
&\frac{ig}{2\sqrt2 \,m_W}\,V_{tb}
[m_b(1+\gamma_5)\tan\beta+m_t(1-\gamma_5)\cot\beta], \nonumber \\
&H^-  t\bar b: &\qquad
&\frac{ig}{2\sqrt2 \,m_W}\,V_{tb}^*
[m_b(1-\gamma_5)\tan\beta+m_t(1+\gamma_5)\cot\beta].
\end{alignat}

The $H_jZZ$ ($H_jW^+W^-$) coupling is, relative to that of the SM, given by
\begin{equation} \label{Eq:ZZH}
H_j ZZ\  (H_jW^+W^-): \qquad 
[\cos\beta R_{j1}+\sin\beta R_{j2}], \quad\text{for }j=1.
\end{equation}
Note that when $H_1$ is CP-odd ($H_1=A$), then $c_2=0$ and this vector
coupling vanishes \cite{Cervero:2012cx}.
Finally, the $H_jH^+W^-$ coupling is given by \cite{ElKaffas:2006nt}
\begin{equation} \label{Eq:HHchW}
H_j H^\pm W^\mp: \qquad 
\frac{g}{2}
[\mp i(\sin\beta R_{j1}-\cos\beta R_{j2})+ R_{j3}]
(p_\mu^j-p_\mu^\pm).
\end{equation}

\section{Constraining the parameter space}
\label{Sec:bounds}
\setcounter{equation}{0}
The multi-dimensional type-II 2HDM parameter space is severely restricted
by a variety of theoretical and experimental constraints, which are
discussed in the following.

\subsection{Theoretical constraints}
\label{subsect:theoconstraints}

We impose three classes of theoretical constraints:
\begin{itemize}
\item
{\bf Positivity}:
In order to have a stable potential, we impose positivity,
$V(\Phi_1,\Phi_2)>0$ as $|\Phi_1|, |\Phi_2| \to\infty$
\cite{Deshpande:1977rw,Nie:1998yn,Kanemura:1999xf}.  Additionally, we must insist on a non-trivial solution to 
Eq.~(\ref{Eq:M_3}): $M_3^2>0$ and $M_2\le M_3$.
While positivity may be satisfied for the given parameters of the potential, the considered minimum need not be the global 
one. However, it has been shown that, if a local charge-conserving minimum exists, then there can be no charge-breaking minimum 
\cite{Ferreira:2004yd,Barroso:2005sm,Barroso:2007rr}.
Nevertheless, the potential of the 2HDM can have more than one charge-conserving minimum. We therefore check that the 
minimum obtained is the global one, following the approach of Ref.~\cite{Grzadkowski:2010au}.

\item
{\bf Tree-level unitarity}:
We also impose tree-level unitarity on Higgs--Higgs scattering
\cite{Kanemura:1993hm,Akeroyd:2000wc,Arhrib:2000is,Ginzburg:2003fe,Ginzburg:2005dt}. These conditions have a rather 
dramatic effect at ``large'' values of $\tan\beta$ and $M_{H^\pm}$, though some tuning of $\mu$ can extend the allowed 
range to larger values of $\tan\beta$ \cite{Aoki:2011wd}.

\item
{\bf Perturbativity}:
We impose the following upper bound on all $\lambda$'s:
\begin{equation} \label{Eq:xi}
|\lambda_i|<4\pi\xi,
\end{equation}
with $\xi=0.8$, meaning $|\lambda_i|\lsim10$. The effect of this is to restrict large values of the
masses, unless the soft parameter $\mu$ is comparable to $M_2$ and $M_{H^\pm}$.
\end{itemize}
For illustrations of how these theory constraints cut into the parameter space, see 
Refs.~\cite{ElKaffas:2006nt,ElKaffas:2007rq}.

\subsection{Experimental constraints}
\label{subsect:expconstraints}

Below, we list the different experimental constraints that are
important. The SM predictions of the flavour observables quoted in
this subsection are obtained using SuperIso v3.2 \cite{Mahmoudi:2007vz,Mahmoudi:2008tp}.
\begin{itemize}
\item {\boldmath $B\to X_s \gamma$}: 
  This rare FCNC inclusive decay receives contributions from the
  charged Higgs boson that can be comparable to the $W^\pm$
  contribution in the SM. Since the charged Higgs state always
  contributes positively to the corresponding BR, it is an effective
  tool to probe the type-II 2HDM.  The most up-to-date SM prediction
  for this decay, at the Next-to-Next-to-Leading Order (NNLO), gives
  \cite{Misiak:2006zs,Misiak:2006ab,Mahmoudi:2007vz,Mahmoudi:2008tp,
Chetyrkin:1996vx,Buras:1997bk,Bobeth:1999mk,Gambino:2001ew,Buras:2002tp,Borzumati:1998tg,Misiak:2004ew,Melnikov:2005bx,Czakon:2006ss,Bauer:1997fe,Neubert:2004dd,Mahmoudi:2009zx}:
\begin{equation}
\rm{BR}(\bar{B} \to X_s \gamma)_{\rm{SM}} = (3.11\pm0.22)\times10^{-4},
\end{equation}
while the combined experimental value from HFAG points to a larger value \cite{Asner:2010qj}:
\begin{equation}
\rm{BR}(\bar{B} \to X_s \gamma)_{\mathrm{exp}}=(3.55\pm0.24\pm0.09)\times10^{-4}.
\end{equation}
For type-II Yukawa interactions, which we consider here, light charged Higgs bosons are excluded by this observable. 
The actual limit is sensitive to higher-order QCD effects and is of
the order of 300~GeV, being more severe at low values of $\tan\beta$ 
\cite{Ciuchini:1997xe,Borzumati:1998tg,WahabElKaffas:2007xd,Mahmoudi:2009zx}.
Recently, a higher-order calculation \cite{Hermann:2012fc} concludes
that the 95\% C.L.\ is at 380~GeV. While adopting the more conservative
limit of 300~GeV, we shall also discuss this more restrictive one.

\item
{\boldmath $B_u\to \tau \nu_\tau$}:
In contrast to the $b \to s\gamma$ transitions, where the charged Higgs state participates in loop diagrams, the process 
$B_u\to\tau\nu_\tau$ can be mediated by $H^\pm$ already at tree level. Since this decay is helicity suppressed in the SM, 
whereas there is no such suppression for spinless $H^\pm$ exchange in the limit of large $\tan\beta$, these two contributions 
can be of similar magnitude \cite{Hou:1992sy}.
The 2HDM contribution factorises in the ratio $R_{B\tau\nu}$ as compared to the SM value. This decay suffers from uncertainties from $f_B$ and $V_{ub}$, and using $f_B=192.8\pm9.9$ MeV \cite{Laiho:2009eu}, and the combined value $|V_{ub}|=(3.92\pm0.46)\times10^{-3}$ 
\cite{Charles:2004jd}, the SM BR evaluates numerically to \cite{Mahmoudi:2007vz,Mahmoudi:2008tp}:
\begin{equation}
\mathrm{BR}(B_u\to\tau\nu_\tau)_\mathrm{SM}=(1.01\pm 0.29)\times 10^{-4}.
\end{equation}
The SM prediction is compared to the current HFAG value\footnote{The latest BaBar result $\mathrm{BR}(B_u\to\tau\nu_\tau)=(1.83^{+0.53}_{-0.49} \pm 0.24)\times 10^{-4}$ \cite{Lees:2012ju} is not included in this average.} \cite{Asner:2010qj}
\begin{equation}
\mathrm{BR}(B_u \to \tau\nu_\tau)_\mathrm{exp}=(1.64\pm 0.34)\times 10^{-4}
\end{equation}
by forming the ratio
\begin{equation}
R_{B\tau\nu}^{\mathrm{exp}}\equiv \frac{\mathrm{BR}(B_u\to\tau\nu_\tau)_{\mathrm{exp}} }{\mathrm{BR}(B_u\to\tau\nu_\tau)_{\mathrm{SM}}} =1.62\pm 0.54.
\label{Rtaunu} 
\end{equation}
In the framework of the 2HDM 
this leads to the exclusion of two sectors of the ratio $\tan\beta/M_{H^\pm}$ \cite{Hou:1992sy,Grossman:1994ax,Grossman:1995yp,Akeroyd:2003zr,Akeroyd:2007eh,Eriksson:2008cx,Mahmoudi:2009zx}.

\item {\boldmath $B\to D\tau \nu_\tau$}: 
  Compared to $B_u\to\tau\nu_\tau$, the semi-leptonic decays $B \to
  D\ell\nu$ have the advantage of depending on $|V_{cb}|$, which is
  known to greater precision than $|V_{ub}|$. In addition, the
  $\mathrm{BR}(B \to D \tau \nu_\tau)$ is about $50$ times larger than
  the $\mathrm{BR}(B _u\to \tau \nu_\tau)$ in the SM. The experimental
  determination remains however very complicated due to the presence
  of at least two neutrinos in the final state.  The ratio
\begin{equation}
\xi_{D\ell\nu_\tau}
=\frac{\text{BR}(B\to D\tau\nu_\tau)}{\text{BR}(B\to D \ell\nu_\ell)},
\end{equation}
where the 2HDM contributes only to the numerator, allows one to reduce some of the theoretical uncertainties.
The SM prediction for this ratio is \cite{Mahmoudi:2007vz,Mahmoudi:2008tp}
\begin{equation}
\xi^{\rm{SM}}_{D\ell\nu}=(29.7\pm 3)\times 10^{-2},
\end{equation}
and the most recent experimental result by the BaBar collaboration is \cite{Lees:2012xj}
\begin{equation}
\xi_{D\ell\nu}^{\mathrm{exp}} = (44.0 \pm 5.8 \pm 4.2) \times 10^{-2}.
\end{equation}
This ratio is also sensitive to a light charged Higgs boson, and leads
to complementary constraints to the ones following from $B_u\to \tau \nu_\tau$
\cite{Grzadkowski:1991kb,Nierste:2008qe,Kamenik:2008tj,Eriksson:2008cx,Mahmoudi:2009zx}.

\item
{\boldmath $D_s\to \tau \nu_\tau$}:
Constraints on a light charged Higgs can be obtained, competitive with those obtained from $B_u\to\tau \nu_\tau$ \cite{Akeroyd:2009tn}. The main uncertainty here is due to the decay constant $f_{D_s}$.
The SM prediction for this decay is \cite{Mahmoudi:2007vz,Mahmoudi:2008tp}:
\begin{equation}
\mathrm{BR}(D_s\to\tau\nu_\tau)_\mathrm{SM}=(5.11\pm 0.13)\times 10^{-2},
\end{equation}
using $f_{D_s} = 248\pm 2.5$ MeV \cite{Davies:2010ip}, and the current world average of the experimental measurements gives \cite{Asner:2010qj}:
\begin{equation}
\mathrm{BR}(D_s \to \tau\nu_\tau)_\mathrm{exp}=(5.38\pm0.32)\times 10^{-2}.
\end{equation}

\item
{\boldmath $B_{d,s}\to \mu^+\mu^-$}:
These decays are helicity suppressed in the SM and can receive sizeable enhancement or depletion from Higgs-mediated 
contributions.  At large $\tan\beta$, the non-observation of these decay modes imposes a
lower bound on the charged Higgs boson mass \cite{Logan:2000iv,Bobeth:2001sq}.
The most stringent limits for their BRs were reported
very recently by the LHCb collaboration \cite{Aaij:2012ac}:
\begin{eqnarray}
&&\mathrm{BR}(B_s \to \mu^+ \mu^-) < 4.5 \times 10^{-9},\\
&&\mathrm{BR}(B_d \to \mu^+ \mu^-) < 1.0 \times 10^{-9},
\end{eqnarray}
at 95\% C.L. Combining these values with the limits from ATLAS and CMS \cite{CMS-PAS-BPH-12-009} results in even stronger bounds:
\begin{eqnarray}
&&\mathrm{BR}(B_s \to \mu^+ \mu^-) < 4.2 \times 10^{-9},\\
&&\mathrm{BR}(B_d \to \mu^+ \mu^-) < 8.1 \times 10^{-10}.
\end{eqnarray}
The SM predictions for these branching ratios are \cite{Mahmoudi:2007vz,Mahmoudi:2008tp,Mahmoudi:2012un}:
\begin{align}
\mathrm{BR}(B_s \to \mu^+ \mu^-) &= (3.53 \pm 0.38) \times 10^{-9},\\
\mathrm{BR}(B_d \to \mu^+ \mu^-) &= (1.1 \pm 0.1) \times 10^{-10},
\end{align}
with BR($B_s \to \mu^+ \mu^-$) being the more constraining. The largest uncertainty is from $f_{B_s}$, we used $f_{B_s} = 234 \pm 10$ MeV for our evaluation.
In the type-II 2HDM, the experimental limits can be reached 
for very large values of the Yukawa couplings and small charged Higgs boson masses. The constraining power of 
$B_{d,s}\to\mu^+\mu^-$ in this study is hence rather limited as compared to the other flavour observables.

\item
{\boldmath $B^0-\bar{B}^0$} {\bf mixing}:
Due to the possibility of $H^\pm$ exchange, in addition to
$W$ exchange, the $B^0-\bar{B}^0$ mixing constraint, which is sensitive to
the term $m_t\cot\beta$ in the Yukawa couplings (\ref{Eq:Yukawa-charged-II}), excludes low values of $\tan\beta$ and low 
values of $M_{H^\pm}$
\cite{Abbott:1979dt,Athanasiu:1985ie,Glashow:1987qe,Geng:1988bq,Inami:1980fz,Urban:1997gw,Mahmoudi:2009zx}.
The non-perturbative decay constant $f_{B_d}$ and the bag parameter $\hat B_d$ which are evaluated simultaneously from 
lattice QCD constitute the largest theoretical uncertainty. 

\item
{\boldmath $R_b$}:
The branching ratio $R_b \equiv \Gamma_{Z\to b\bar b} /\Gamma_{Z\to {\rm had}}$ would also be affected by Higgs boson exchange.
The contributions from neutral Higgs bosons to $R_b$ are negligible \cite{ElKaffas:2006nt}, however, charged Higgs boson 
contributions, via the $H^\pm bt$ Yukawa coupling, as
given by \cite{Denner:1991ie}, Eq.~(4.2), exclude low values of $\tan\beta$ and low $M_{H^\pm}$.  


\item
{\boldmath{$pp \to H_jX$:}}
Two aspects of the recent neutral Higgs searches at the LHC are
considered \cite{ATLAS:2012ae,Chatrchyan:2012tx}: 
\begin{itemize}
\item
The production and subsequent decay of a neutral Higgs to
$\gamma\gamma$, around $M=125~\text{GeV}$ is taken to be within a
factor of 2 from the SM rate. Assuming the dominant production to be via gluon fusion, this can be approximated as 
$0.5\leq R_{\gamma\gamma}\leq2$, where we define
\begin{equation} \label{Eq:R_gammagamma}
R_{\gamma\gamma}=\frac{\Gamma(H_1\to gg){\rm BR}(H_1\to\gamma\gamma)}
{\Gamma(H_\text{SM}\to gg){\rm BR}(H_\text{SM}\to\gamma\gamma)}.
\end{equation}
We take into account (1) the modified scalar coupling to the fermion or
$W$ in the loop, (2) the pseudoscalar contribution, and (3) the charged
Higgs contribution on the $\gamma\gamma$ side.
This condition (\ref{Eq:R_gammagamma}) mainly constrains the Yukawa couplings of $H_1$. In
particular, the (dominant) $H_1t\bar t$ contribution to the loop integrals should be comparable to that of the
SM, meaning
\begin{equation}
\frac{s_1^2c_2^2}{\sin^2\beta}+\frac{s_2^2}{\tan^2\beta}P^2(\tau_t)={\cal O}(1),
\end{equation}
where $P(\tau)$ represents the ratio of the pseudoscalar and the
scalar contributions to the loop integral \cite{Djouadi:2005gj}, with $\tau_t=M_1^2/(4m_t^2)$.
At low $\tan\beta$ there is some freedom, either $s_1^2c_2^2$ or $s_2^2$ should be of
order unity, whereas at high $\tan\beta$
this constraint requires $\alpha_1\simeq\pm\pi/2$ and
$\alpha_2\simeq0$.
\item
The production and subsequent decay, dominantly via $ZZ$ and $WW$, is constrained in the mass range 
$130~\text{GeV}\lsim M \lsim600~\text{GeV}$. We consider the quantity
\begin{equation} \label{Eq:R_ZZ}
R_{ZZ}=\frac{\Gamma(H_j\to gg){\rm BR}(H_j\to ZZ)}
{\Gamma(H_\text{SM}\to gg){\rm BR}(H_\text{SM}\to ZZ)},
\end{equation}
for $j=2,3$ and require it to be below the stronger 95\% CL obtained
by ATLAS or CMS. This constraint thus affects the product of the Yukawa
and gauge couplings of $H_2$ and $H_3$ (see Eqs.~(\ref{Eq:H_j_Yuk})
and (\ref{Eq:ZZH})). In the limit of unchanged total width, this implies
\begin{equation} \label{Eq:ZZconstraint}
\biggl[\frac{R_{j2}^2}{\sin^2\beta}+\frac{R_{j3}^2}{\tan^2\beta}\biggr]
[\cos\beta R_{j1}+\sin\beta R_{j2}]^2<\eta, \quad j=2,3,
\end{equation}
where $\eta$ is the 95\% CL on $\sigma/\sigma_\text{SM}$.
For $\tan\beta$ of the order of unity, the first factor is ``small''
when $R_{j1}^2$ is of order unity, whereas the second factor is
``small'' when $|R_{j3}|$ is of order unity.

For larger values of $\tan\beta$, we may substitute $\alpha_1\simeq\pm\pi/2$ and $\alpha_2\simeq0$ from the
above consideration, whereupon the constraint (\ref{Eq:ZZconstraint}) takes the form:
\begin{equation}
\frac{\cos^2\beta}{\tan^2\beta}(s_3c_3)^2<\eta,
\end{equation}
which is not very strong. In particular, it is automatically satisfied
at large $\tan\beta$.

For the total widths of $H_2$ and $H_3$ entering in the numerator of
(\ref{Eq:R_ZZ}), we include also $H_j\to H_1H_1$ and $H_j\to
H_1Z$, in addition to the familiar decay modes of $H_1$. The relevant
couplings can be found in \cite{Osland:2008aw} and \cite{ElKaffas:2006nt}.
\end{itemize}

\item
{\boldmath $T$ and $S$:}
For the electroweak ``precision observables'' $T$ and $S$, we impose the bounds $|\Delta T|<0.10$,
$|\Delta S|<0.10$ \cite{Nakamura:2010zzi}, at the 1-$\sigma$ level,
within the framework of
Refs.~\cite{Grimus:2007if,Grimus:2008nb}. 
While $S$ is not very restrictive,
$T$ gets a positive contribution from a splitting between the masses of charged and neutral Higgs bosons, whereas a pair 
of neutral ones give a negative contribution.

\item
{\bf Electron Electric Dipole Moment (EDM)}:
The bound on the electron EDM constrains the allowed amount of CP
violation of the model. We adopt the bound \cite{Regan:2002ta} (see also
\cite{Pilaftsis:2002fe}):
\begin{equation}
|d_e|\lsim1\times10^{-27} [e\,\text{cm}],
\end{equation}
at the 1-$\sigma$ level.
The contribution due to neutral Higgs exchange, via the
two-loop Barr--Zee effect \cite{Barr:1990vd}, is given by Eq.~(3.2) of \cite{Pilaftsis:2002fe}.

\end{itemize}

In contrast to the MSSM, in the 2HDM, an additional contribution to the muon anomalous magnetic moment arises only at the 
two-loop level.
Since we are considering heavy Higgs bosons ($M_1, M_{H^\pm} \gsim 100~\text{GeV}$) therefore, according to
\cite{Cheung:2003pw,Chang:2000ii,WahabElKaffas:2007xd}, the 2HDM
contribution to the muon anomalous magnetic moment is negligible even for $\tan\beta$ as large as $\sim 40$.

The above constraints are not independent. Therefore, we do not
attempt to add their contributions to an overall $\chi^2$, but rather require that none of them should be violated by more 
than $2\sigma$. The LHC constraints are imposed at the quoted 95\% C.L.

Among these constraints, basically $b\to s\gamma$ requires
$M_{H^\pm}\gsim 300$ $(380)~\text{GeV}$, while the 
different $B$-meson constraints impose additional constraints at low and high values of $\tan\beta$ (the latter are 
basically excluded anyway, by the unitarity constraints). The $T$ constraint prevents the masses of the neutral and 
charged Higgs bosons from being very different, and thus effectively provides a cut-off at high masses.

\subsection{Two scenarios}

In order to develop some intuition for the viable parts of the
parameter space, we shall here consider two scenarios. In both of them,
we take the lightest neutral Higgs boson mass to be $M_1=125~\text{GeV}$. Furthermore:
\begin{itemize}
\item
{\bf Scenario 1: low $\tan\beta$, intermediate different masses $M_2$ and
  $M_{H^\pm}$ (non-decoupling regime):} 
\begin{alignat}{2}
\tan\beta=1,\, 2,\,3; \quad
M_1&=125~\text{GeV}, &\quad
M_2&=150,200,300,400,500~\text{GeV}, \nonumber \\
M_{H^\pm}&=300-600~\text{GeV}, &\quad
\mu&=200~\text{GeV}.
\end{alignat}
\item
{\bf Scenario 2: high $\tan\beta$, heavy degenerate masses $M_2$ and  $M_{H^\pm}$ (decoupling regime):} 
\begin{alignat}{2}
\tan\beta=5,10,20; \quad
M_1&=125~\text{GeV}, &\quad
M_2&=400,\ 600~\text{GeV}, \nonumber \\
M_{H^\pm}&\simeq M_2, &\quad
\mu&\simeq M_2.
\end{alignat}
\end{itemize}

For scenario~1, we will consider a range of charged Higgs boson masses, from 300 to 600~GeV (for 700~GeV, only a few viable parameter points are
found), and a range of $\tan\beta$ values. We will typically consider
small values of $\tan\beta$, of the order of 1. High values lead (for
fixed $\mu$) to
conflict with the unitarity constraints.

Scenario 2 is rather fine-tuned. The masses $M_2$ and $M_{H^\pm}$ have to be very close to
$\mu$, in order to avoid conflict with the unitarity constraints. It
is discussed in some detail in the appendix. There, it is
shown that in addition to the SM-like region of
$\alpha_1\simeq\pm\pi/2$ with $\alpha_2\simeq0$ ($H_2$ or $H_3$ being
odd under CP), there is also another
region, with $\alpha_1\simeq0$ and $\alpha_2\simeq\pm\pi/2$ (with $H_1$
being odd, see \cite{Gerard:2007kn,Dermisek:2010mg}).

\subsection{Studies in $\alpha$ space}

\subsubsection{Scenario 1}
\label{subsect:scenario1}

\begin{figure}[htb] 
 \includegraphics[angle=0,width=0.9\textwidth
  ]{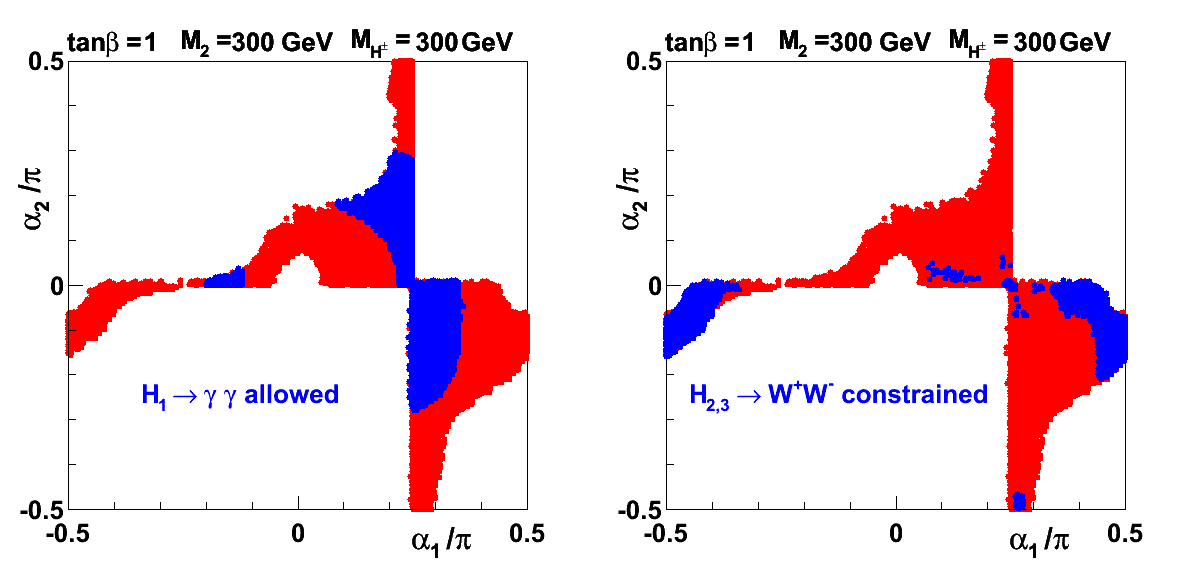}
\caption{Allowed regions in the $\alpha_1$--$\alpha_2$
  parameter space, without (red) and with (blue) LHC constraints, for
  $M_2=300~\text{GeV}$ and the additional parameters
  given in Eq.~(\ref{Eq:plot-params}).
Left, blue: regions surviving the $H_1\to\gamma\gamma$
constraint.
Right, blue: regions surviving the $H_{2,3}\to W^+W^-$ constraint.
\label{Fig:alpha12-1-300-1001}}
\end{figure}

We shall first consider separately the LHC constraints on $H_1\to\gamma\gamma$
(an allowed range) and $H_{2,3}\to W^+W^-$ (an upper bound).
For fixed additional input parameters,
\begin{equation} \label{Eq:plot-params}
M_{H^\pm}=300,\ 500~\text{GeV}, \quad
\tan\beta=1, \quad
\mu=200~\text{GeV},
\end{equation}
we show in fig.~\ref{Fig:alpha12-1-300-1001} the result of a scan
over $10^6$ sets of ($\alpha_1,\alpha_2,\alpha_3$). The blue regions satisfy all
the constraints discussed above, as well as one of these LHC
constraints, whereas the red regions do not satisfy the LHC constraint
considered, but all other constraints discussed in sections~\ref{subsect:theoconstraints}
and \ref{subsect:expconstraints}.

The allowed regions are for these parameters rather independent of
$M_{H^\pm}$ in the range 300--400~GeV,
but start shrinking around 500~GeV and vanish around $M_{H^\pm}\sim600~\text{GeV}$. (If we allow $\xi=1$,
see Eq.~(\ref{Eq:xi}), the allowed values of $M_{H^\pm}$ reach out to
about 700~GeV.) 
The underlying checkered pattern is due to the
positivity constraint, together with $M_3^2\geq M_2^2$.

When we impose the LHC constraints discussed above \cite{ATLAS:2012ae,Chatrchyan:2012tx}
 (scanning now over $5\cdot10^6$ points), we obtain the blue regions for
the effects of
the $H_1\to\gamma\gamma$ (left) and the $H_{2,3}\to W^+W^-$ (right)
constraint.
For the considered case of $\tan\beta=1$,
$M_2=M_{H^\pm}=300~\text{GeV}$ we see that these only very marginally overlap.

\begin{figure}[htb] 
 \includegraphics[angle=0,width=0.9\textwidth
  ]{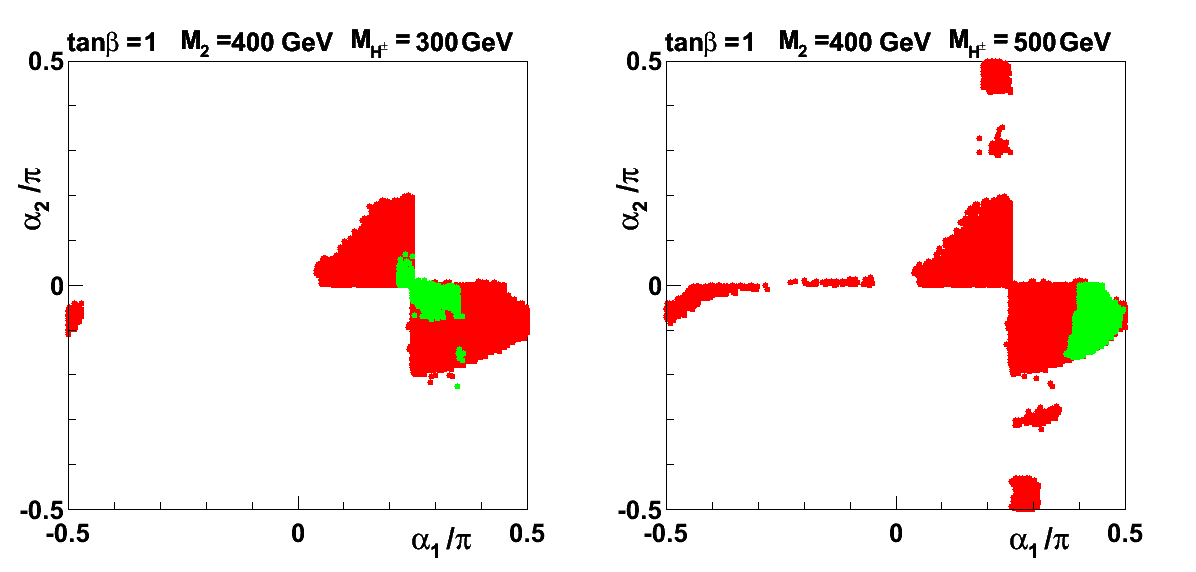}
\caption{Allowed regions in the $\alpha_1$--$\alpha_2$
  parameter space, without (red) and with (green) the LHC constraints, for
  $M_2=400~\text{GeV}$ and the additional parameters
  given in Eq.~(\ref{Eq:plot-params}).
\label{Fig:alpha12-1-400}}
\end{figure}

Imposing then both of these LHC constraints, we show in
figure~\ref{Fig:alpha12-1-400} the resulting surviving parameter space
in green on top of the red regions, now for $M_2=400~\text{GeV}$, and
two values of charged-Higgs mass, 300 and 500~GeV.
Note that although the LHC experiments exclude an SM Higgs
with a mass from about 130~GeV to about 600~GeV, there are still viable regions of parameter space for the
second (and third) Higgs state to be in this region, since it may couple more weakly
than the SM Higgs boson.

A striking first observation is that the allowed regions are very much
reduced, only values close to $\alpha_1=\pm\pi/4$ are now
allowed for the low value of $M_{H^\pm}$, and a somewhat higher value
for the higher value of $M_{H^\pm}$. This is a little different from the results reported recently for the CP-conserving model, where it was
found that only a region around $\alpha=0$ is allowed \cite{Ferreira:2011aa}. We recall that
in the particular CP-conserving limit of $\alpha_2=\alpha_3=0$
(corresponding to the heaviest one, $H_3$, being odd under CP),
$\alpha=0$ corresponds to $\alpha_1=\pm\pi/2$. That region is here found to
violate unitarity.

\begin{figure}[htb] 
\includegraphics[angle=0,width=0.9\textwidth
  ]{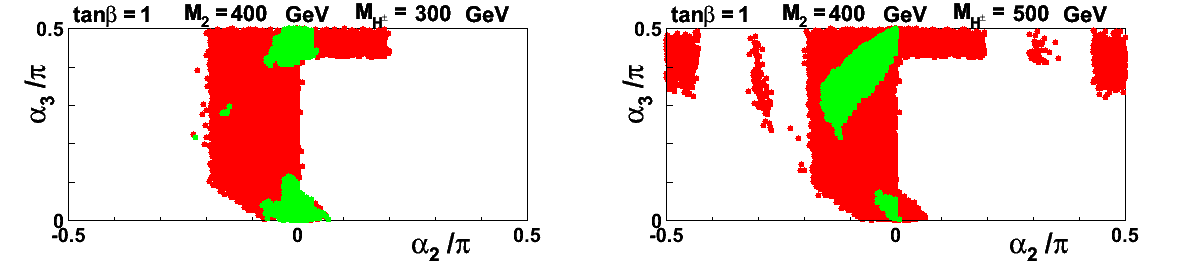}
 \caption{Allowed regions in the $\alpha_2$--$\alpha_3$
  parameter space, without (red) and with (green) the LHC constraints, for the parameters
  given in Eq.~(\ref{Eq:plot-params}).
\label{Fig:alpha23-1-400}}
\end{figure}

For a complementary view of the allowed region, we show in
fig.~\ref{Fig:alpha23-1-400} the corresponding projections onto the
$\alpha_2$--$\alpha_3$ plane. 
Much of the region with $\alpha_2<0$
is populated, whereas most of the region with $\alpha_2>0$ is
excluded. All values of $\alpha_3$ are represented.
When we impose the LHC constraints, the main characteristic is that $\alpha_2$
becomes more restricted, as was also seen in
fig.~\ref{Fig:alpha12-1-400}, and by Barroso et al \cite{Barroso:2012wz}.
Two regions of $\alpha_3$-values are represented: values
close to 0 or $\pi/2$ (in both cases for $\alpha_2$ close to zero).

A comparison with Arhrib et al \cite{Arhrib:2012yv} indicates that we
find a more constrained region, presumably because of our tighter
constraint on $R_{\gamma\gamma}$. In addition, we find that some
points are excluded because of conflict with the electron EDM.

For higher values of $\tan\beta$, several things change.
At some point, also negative values of $\alpha_1$ become allowed (not shown), and
the allowed ranges of $\alpha_1$ move towards $\pm\pi/2$ as
$\tan\beta$ increases. At $\tan\beta=2$ and 3, the allowed region has shrunk by a factor of
2 to 3, compared to $\tan\beta=1$.

For lower values of $M_2$, the impact of the LHC constraints is more
severe, but allowed points are found, for example also for
$M_2=150~\text{GeV}$. For higher values of $M_2$, the allowed region
is restricted to some neighborhood of $\alpha_1=\pi/4$ and
$\alpha_2=0$, with $\alpha_3$ close to 0 or $\pi/2$.

In view of the results shown in
fig.~\ref{Fig:alpha12-1-400}, let us comment on the
special limit
\begin{equation}\label{Eq:special-limit}
\alpha_1=\pi/4, \quad \alpha_2~\text{small} ~(s_2\simeq0).
\end{equation}
Then, the rotation matrix can be simplified as
\begin{equation}
R=
\frac{1}{\sqrt{2}}\begin{pmatrix}
1 & 1 & 0 \\
-c_3 & c_3 & \sqrt{2}s_3 \\
s_3 & -s_3 &\sqrt{2} c_3
\end{pmatrix},
\end{equation}
and the physical states are related to the ``weak'' states $\eta_j$ as
\begin{subequations}
\begin{align}
H_1&=\frac{1}{\sqrt{2}}\bigl(\eta_1+\eta_2\bigr), \\
H_2&=\frac{1}{\sqrt{2}}\bigl(-c_3\eta_1+c_3\eta_2+\sqrt{2} s_3\,\eta_3\bigr), \\
H_3&=\frac{1}{\sqrt{2}}\bigl(s_3\eta_1-s_3\eta_2+\sqrt{2} c_3\,\eta_3\bigr).
\end{align}
\end{subequations}
We recall that, with type-II Yukawa couplings, $\eta_1$ couples to down-type quarks whereas $\eta_2$ couples to up-type 
quarks. Thus, in the limit (\ref{Eq:special-limit}),
the lightest neutral Higgs boson couples coherently to the $b$- and the
$t$-quarks (with strengths proportional to $1/\cos\beta$ and
$1/\sin\beta$, respectively, and thus,  for
$\tan\beta=1$, with the same strength). 
The heavier Higgs bosons, however, have a CP-even content given by
$\eta_1-\eta_2$ (note the minus sign). For $\tan\beta=1$ only the pseudoscalar component,
$\eta_3$, of these heavier Higgs bosons will couple to fermions.
We note that two CP-conserving limits are contained in this scenario:
$H_3=A$ for $\alpha_3=0$ and $H_2=A$ for $\alpha_3=\pi/2$.

\begin{figure}[!htb] 
 \includegraphics[angle=0,width=0.9\textwidth
  ]{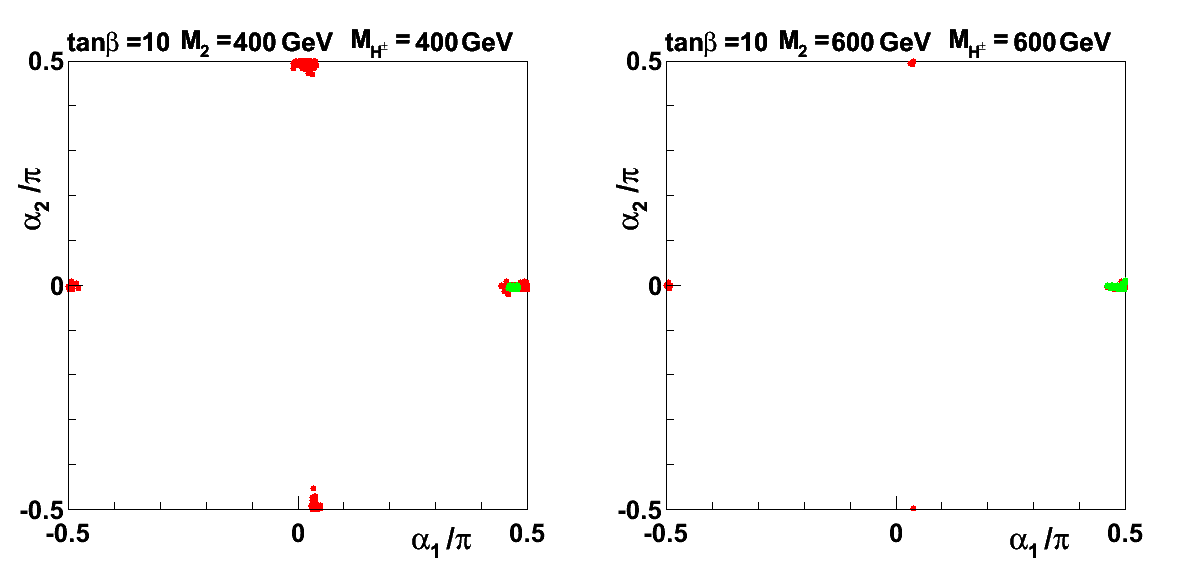}
\caption{Allowed regions in the $\alpha_1$--$\alpha_2$
  parameter space, without (red) and with (green) the LHC constraints,
  for $M_1=125~\text{GeV}$, 
  $M_2=400~\text{GeV}$ (left), $M_2=600~\text{GeV}$ (right) and
$M_2-10~\text{GeV}\leq M_{H^\pm}\leq M_2+10~\text{GeV}$.
\label{Fig:alpha12-hitanb}}
\end{figure}

\subsubsection{Scenario 2}

Again, we start out with an overview of the allowed regions of parameter space in the
absence of LHC constraints. This is presented in red, in
fig.~\ref{Fig:alpha12-hitanb}, for $\tan\beta=10$ and two values of $M_2$, namely 400 (left)
and 600 (right)~GeV. The populated regions are at
$(\alpha_1,\alpha_2)\simeq(\pm\pi/2,0)$
and $\simeq(0,\pm\pi/2)$, the decoupling regions which are discussed
in the appendix.
They are seen to shrink considerably as the masses are increased from 400 to 600~GeV.

When we impose the LHC constraints, the regions near $(\alpha_1,\alpha_2)=(0,
\pm\pi/2)$ are no longer allowed. Neither is the region near
$(\alpha_1,\alpha_2) \simeq (-\pi/2,0)$. This is a CP-conserving
limit, also commented on in section~\ref{subsect:cp-conservation}. Only a small region
near $(\alpha_1,\alpha_2) \simeq (\pi/2,0)$ remains.
The figure includes a range of values for $M_{H^\pm}$ within $[M_2-10~\text{GeV},
M_2+10~\text{GeV}]$. If we take $M_{H^\pm}=M_2$, then
$\alpha_2$ has to be slightly different from zero, meaning
that CP is violated. At the CP-conserving parameter point ($\alpha_1,\alpha_2,\alpha_3$) =
($\pi/2,0,0$), unitarity is violated.

\subsubsection{CP-conserving limits}
\label{subsect:cp-conservation}
\begin{figure}[!htb] 
 \includegraphics[angle=0,width=0.9\textwidth
  ]{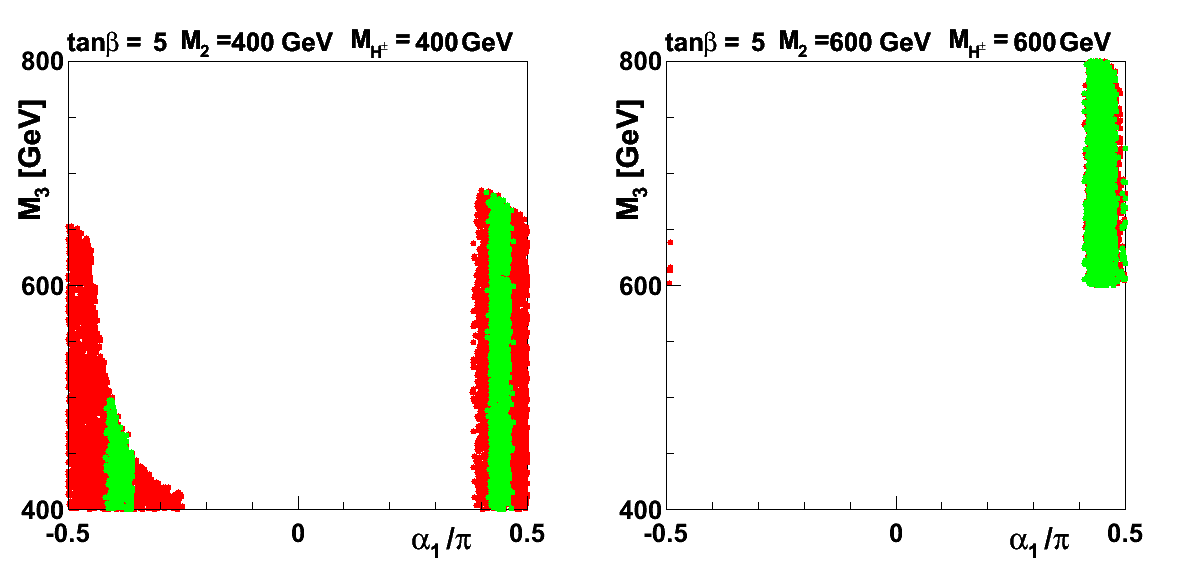}
\caption{CP conserving case. Allowed regions in the $\alpha_1$--$M_3$
  parameter space, without (red) and with (green) the LHC constraints,
  for $M_1=125~\text{GeV}$, 
  $M_2=M_{H^\pm} =400~\text{GeV}$ (left), $M_2=M_{H^\pm} =600~\text{GeV}$ (right).
\label{Fig:alpha1-mh3}}
\end{figure}

Solutions also exist in CP-conserving limits.
With $H_3$ CP-odd ($H_3=A$), we have performed scans at $\tan\beta=3$, 5,
10 and 20, for two heavy-mass cases: (i)
$H_2=M_{H^\pm}\equiv M=400~\text{GeV}$ and (ii)
$M=600~\text{GeV}$.
In the absence of the LHC constraint, some range in $\alpha_1$ around
$\pm\pi/2$ (corresponding to $\alpha$ around 0) is populated. Imposing
the LHC constraints, the range in $\alpha_1$ is constrained, see fig.~\ref{Fig:alpha1-mh3} for the case
$\tan\beta=5$. This is consistent with the allowed regions discussed
in section~\ref{subsect:scenario1}, for
$\alpha_2\simeq\alpha_3\simeq0$ (see figures~\ref{Fig:alpha12-1-400} and \ref{Fig:alpha23-1-400}).
For increasing values of $\tan\beta$ (10 and 20), the allowed bands
move towards the edges and become more narrow (compare
fig.~\ref{Fig:alpha12-hitanb}).

When $H_2$ is CP-odd ($H_2=A$) and $\tan\beta=5$, we find allowed
solutions at $M=400~\text{GeV}$ and 600~GeV. But compared to the case
$H_3=A$, the mass range for $M_3$ is more
constrained.

We do not find any solution when $H_1$ is CP-odd ($H_1=A$). The crucial
LHC constraint is the quantity $R_{\gamma\gamma}$. Others
\cite{Burdman:2011ki,Cervero:2012cx} have argued for an interpretation of the 125~GeV excess in
terms of a pseudoscalar, but we are not able to confirm this. 
We find points which satisfy all other constraints, but not the LHC
constraints. Another recent study agrees with this \cite{Coleppa:2012eh}.

\subsection{Studies in $\tan\beta$--$M_{H^\pm}$ space}

\subsubsection{The unitarity constraint}

The unitarity constraint plays an important role in delimiting high
values of both $\tan\beta$ and $M_{H^\pm}$. This constraint 
requires the $\lambda$'s to be small, which to some extent is achieved
by taking the ``soft'' mass parameter $\mu$ large. In fact, in the
co-called decoupling limit, discussed for the CP-conserving case in
\cite{Gunion:2002zf}, and for the present case in the appendix,
one can respect the unitarity constraints for large masses, provided
$\mu$ is tuned to these masses:
\begin{equation}
M_2\sim M_3\sim M_{H^\pm}\sim\mu.
\end{equation}
For moderate values of $\tan\beta$ ($3-5$), that limit also requires
\begin{equation}
\beta\sim\alpha_1, \quad\alpha_2\sim0, \quad\alpha_3 \text{ arbitrary}.
\end{equation} 
For large values of
$\tan\beta$  ($\gsim5$), this evolves into the region
$(\alpha_1,\alpha_2)\sim(\pi/2,0)$.
Furthermore, an additional region opens up for large masses and large
$\tan\beta$, leading to
\begin{subequations}
\begin{alignat}{2}
&\text{Decoupling 1:} &\quad
(\alpha_1,\alpha_2)&\sim(\pm\pi/2,0), \\
&\text{Decoupling 2:} &\quad
(\alpha_1,\alpha_2)&\sim(0,\pm\pi/2),
\end{alignat} 
\end{subequations}
with $\alpha_3$ arbitrary. Two comments are here in order: (i) because
of the periodicity of the trigonometric functions, regions at
$\alpha_i\simeq-\pi/2$
and $\alpha_i\simeq+\pi/2$
are connected; (ii) the SM limit requires
$\alpha_1\sim\beta$, and is thus contained in the region ``Decoupling 1''.

\begin{figure}[!htb] 
\includegraphics[angle=0,width=0.7\textwidth
  ]{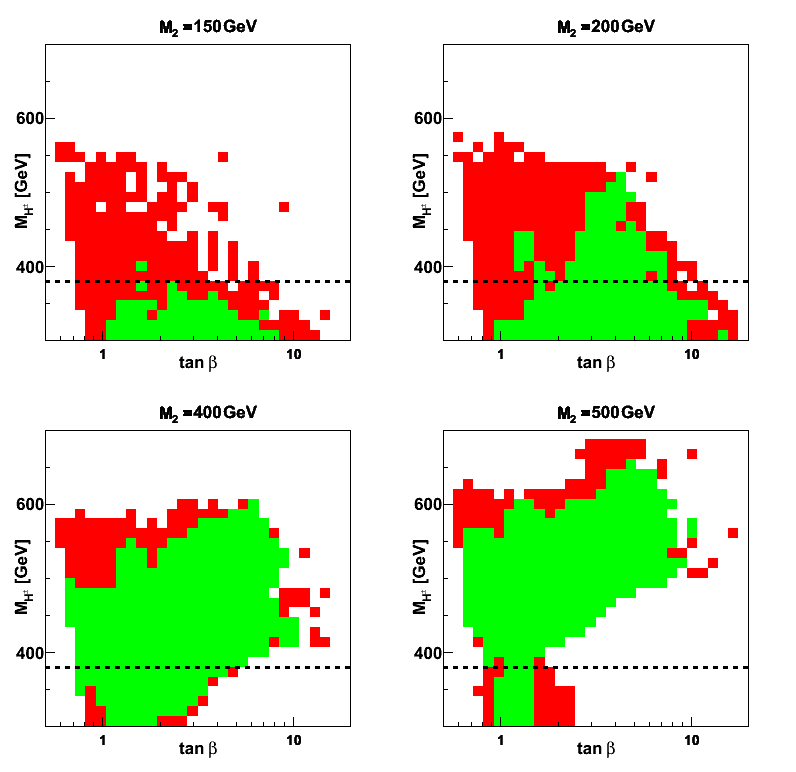}
 \caption{Allowed regions in the $\tan\beta$--$M_{H^\pm}$
  parameter space, without (red) and with (green) the LHC constraints, for $M_1=125~\text{GeV}$
  and four values of $M_2$,
  as indicated.
The dashed lines show the recent bound at 380~GeV \cite{Hermann:2012fc}.
\label{Fig:tanbeta-mh_ch}}
\end{figure}

In view of the above discussion, in order to
determine the maximally allowed ranges of $\tan\beta$ and $M_{H^\pm}$, we scan
over some range in $\mu$, starting at the geometric mean
\begin{equation}
\mu_0=\sqrt{M_2M_{H^\pm}}.
\end{equation}

\subsubsection{The experimental constraints}

In fig.~\ref{Fig:tanbeta-mh_ch} we show allowed regions in
the $\tan\beta$--$M_{H^\pm}$ plane. Again, the larger red region is
allowed in the absence of recent LHC results, whereas the green
region shows what remains compatible with these data. We note some
reduction in the range of charged Higgs masses. Also, at high
$\tan\beta$, the masses $M_2$ and $M_{H^\pm}$ tend to be close, as
discussed above.

\begin{figure}[!htb] 
\includegraphics[angle=0,width=0.7\textwidth
  ]{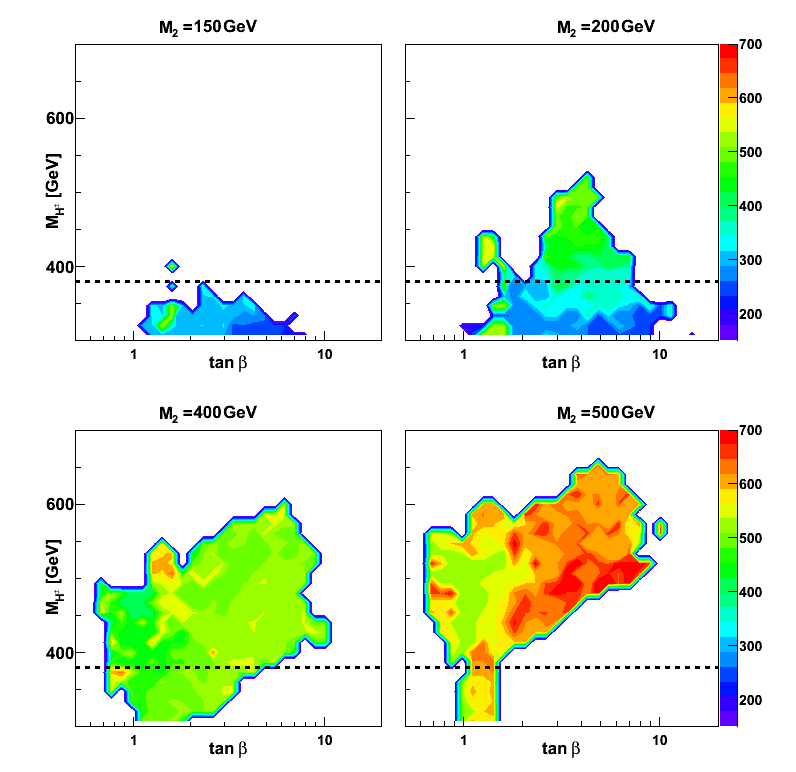}
 \caption{Heaviest neutral Higgs mass, $M_3$ in the $\tan\beta$--$M_{H^\pm}$
  parameter space, with the LHC constraints, for $M_1=125~\text{GeV}$
  and four values of $M_2$,
  as indicated.
The dashed lines show the recent bound at 380~GeV \cite{Hermann:2012fc}.
\label{Fig:mh3-tanbeta-mh_ch-lhc}}
\end{figure}

The ``fractal'' appearance of these plots is in part due to the finite
number of points in the scans. Some could also reflect genuine
``islands'' in parameter space.

In fig.~\ref{Fig:mh3-tanbeta-mh_ch-lhc} we show typical values of
$M_3$. Note that for each point in the allowed part of this plane,
some ranges of $\alpha$'s are allowed (see the previous
subsection). Each set of $\alpha$'s corresponds to a particular value
of $M_3$. The values plotted here are those first encountered in a
random scan over $\alpha$'s. We see that as $M_2$ and $M_{H^\pm}$ increase, also
typical values of $M_3$ increase.

By allowing a larger value of the perturbativity cut-off $\xi$ of
Eq.~(\ref{Eq:xi}), higher masses of $M_{H^\pm}$ would be allowed. For
example, $\xi=1$ permits masses above 600~GeV.
Also the unitarity and the electroweak parameter $T$
constrain this high-mass region. Which of these gives the strongest
limits depends on the other parameters.


\section{Benchmark analysis}
\label{Sec:Results}
\setcounter{equation}{0}
In this section we study the profile of the charged Higgs boson at the LHC in view of the allowed parameter space analysis.
For this, we start by studying a set of candidate benchmark points $P_i$ that allow us to synthesise the main features of the surviving models. 

As we have shown in the previous sections, this model depends on eight parameters. However, since the $\mu$ parameter does not directly participate in the phenomenology of interest, if not specified otherwise, we will consider $\mu=200$ GeV hereafter. The exception will be the high-$\tan\beta$ case, where $\mu$ has to be carefully tuned in order to respect the unitarity constraint. Then, we consider points that pass the constraints, at least in the charged Higgs mass range $300-600$ GeV. We are then left with five parameters: $\tan\beta$, $\alpha_1$, $\alpha_2$, $\alpha_3$ and $M_2$.

\begin{table}[ht]
\begin{center}
\begin{tabular}{|c|c|c|c|c|c|c|}
\hline
\ & $\alpha_1/\pi$ & $\alpha_2/\pi$ & $\alpha_3/\pi$ & $\tan\beta$ & $M_2$ & $M_{H^\pm}^\text{min} ,M_{H^\pm}^\text{max}$ \\
\hline
 $P_1$ & $0.23$ & $0.06$ & $0.005$ & $1$ & $300$ & 300,325 \\
 $P_2$ & $0.35$ & $-0.014$ & $0.48$  & $1$ & $300$ & 300,415\\
 $P_3$ & $0.35$ & $-0.015$ & $0.496$ & $1$ & $350$ & 300,450\\
 $P_4$ & $0.35$ & $-0.056$ & $0.43$ & $1$ & $400$ & 300,455\\
 $P_5$ & $0.33$ & $-0.21$ & $0.23$ & $1$ & $450$ & 300,470\\
 $P_6$ & $0.27$ & $-0.26$ & $0.25$ & $1$ & $500$ & 300,340\\
\hline
 $P_7$ & $0.39$ & $-0.07$ & $0.33$ & $2$ & $300$ & 300,405 \\
 $P_8$ & $0.34$ & $-0.03$ & $0.11$ & $2$ & $400$ & 300,315\\
\hline
 $P_9$ & $0.47$ & $ -0.006$ & $0.05$ & $10$ & $400$ & 400,440 \\
 $P_{10}$ & $0.49$ & $-0.002$ & $0.06$ & $10$ & $600$ & 600,700\\
\hline
\end{tabular}
\end{center}
\caption{Benchmark points selected from the allowed parameter space. Masses $M_2$ and allowed range of $M_{H^\pm}$ are in GeV. For $P_1$--$P_8$, $\mu=200~\text{GeV}$, whereas for $P_9$ and $P_{10}$, $\mu=M_2$.  \label{points}}
\end{table}

In table~\ref{points}, we list a set of 10 candidate benchmark points: we consider the most illustrative four of them for determining cross sections and relevant BRs, the rest will be discussed only qualitatively.

This set of points has distinctive characteristics in the phenomenology, as will be shortly made clear.
Also, we note that $P_{9}$ and $P_{10}$ are nearly CP-conserving, with $\alpha_2\simeq\alpha_3\simeq0$. All the others correspond to CP-violating scenarios.

The model has been implemented through the LanHEP module \cite{Semenov:1996es} (see \cite{Maderetal} for details) and the following analysis has been performed by means of the CalcHEP package \cite{Pukhov:2004ca,calchep_man}. Furthermore, we have used the CTEQ6.6M \cite{Nadolsky:2008zw} set of five-flavour parton distribution functions (PDFs). Due to their relevance at hadron colliders, the effective $ggH_i$, $\gamma\gamma H_i$ and $\gamma ZH_i$ vertices have been implemented by means of a link between CalcHEP and LoopTools \cite{Hahn:1998yk}, and the numerical results have been cross-checked against the analytical results in \cite{Posch:2010hx}.

\begin{figure}[htb]
  \includegraphics[angle=0,width=0.48\textwidth ]{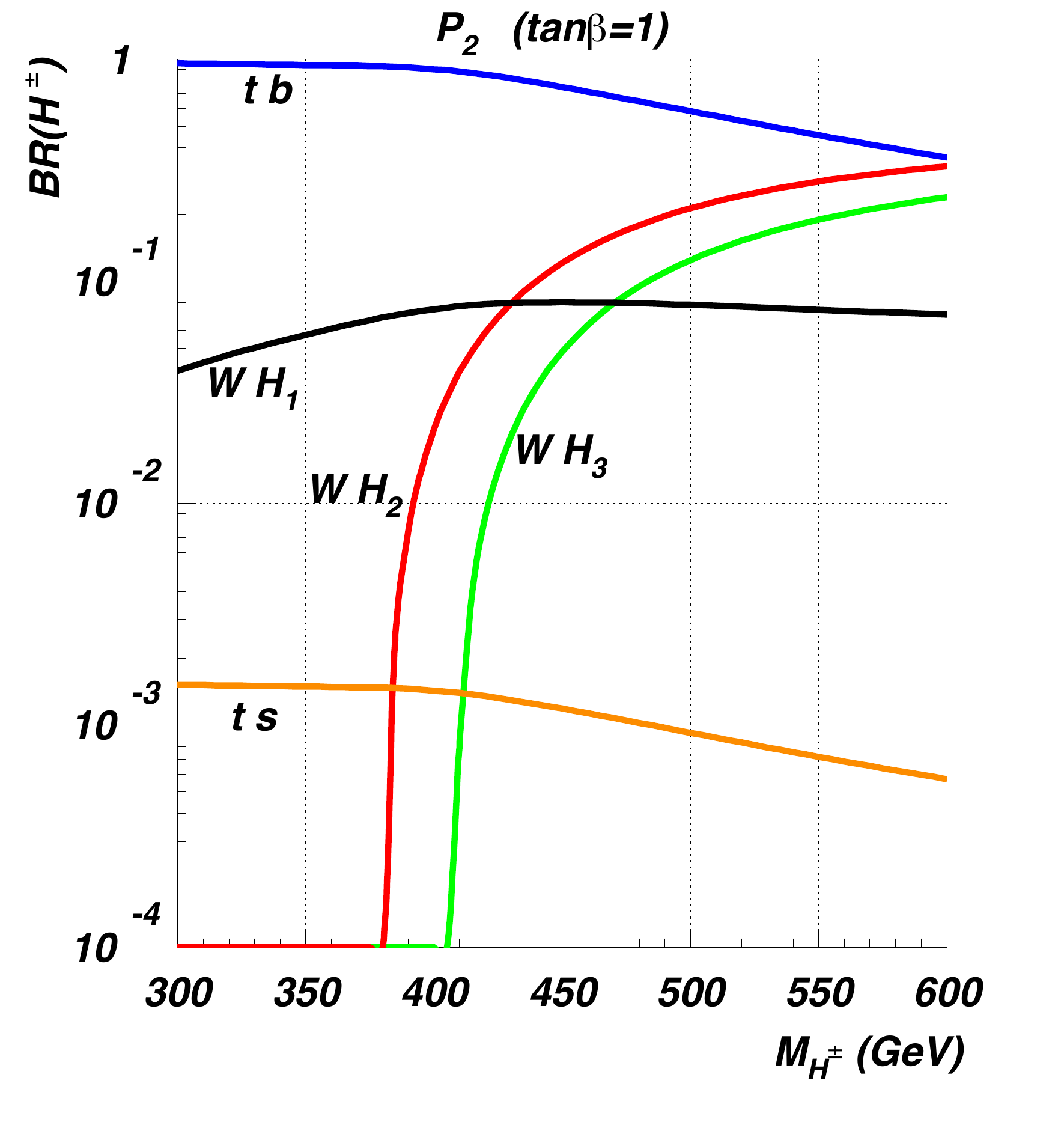}
  \includegraphics[angle=0,width=0.48\textwidth ]{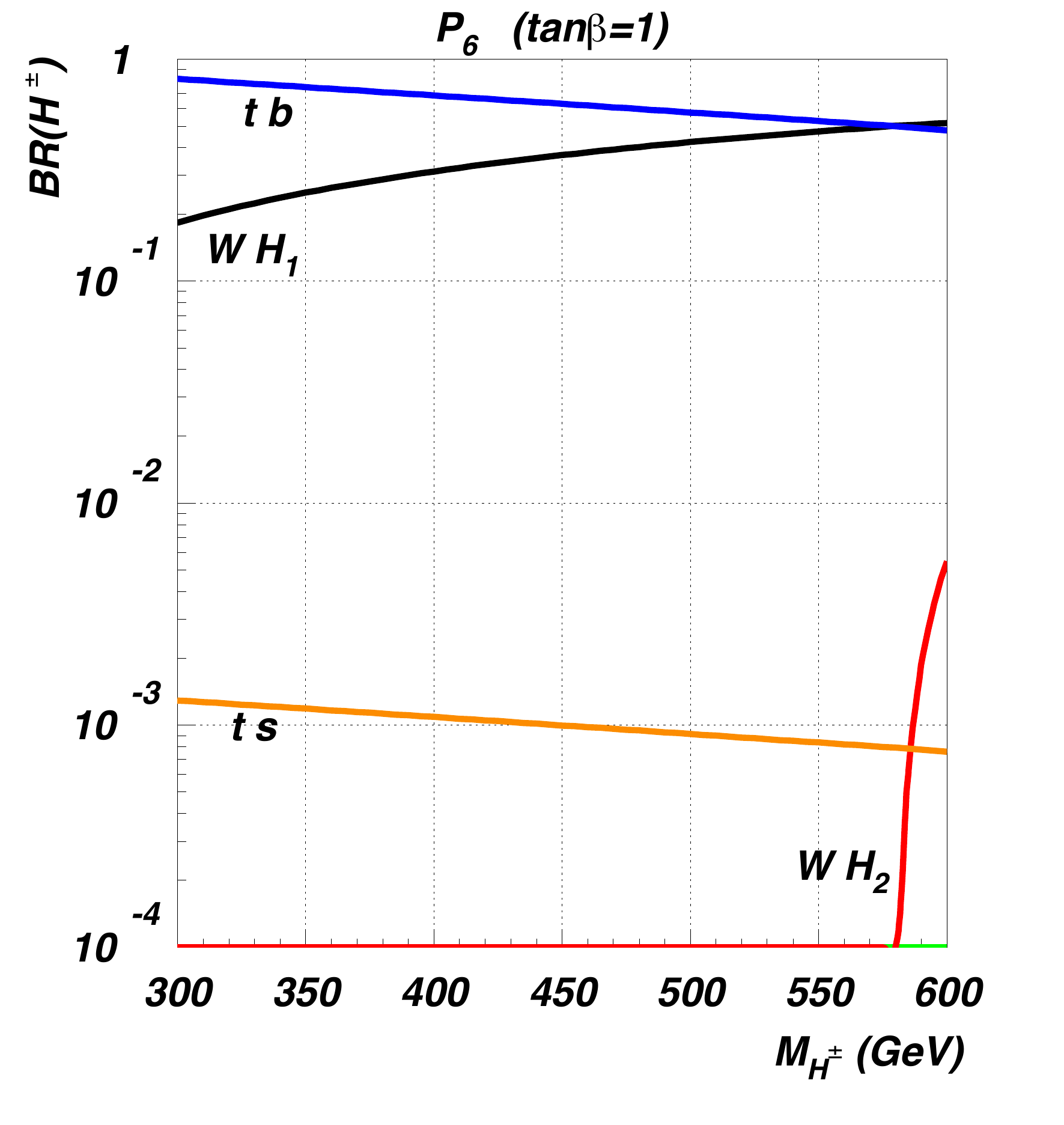} \\
  \includegraphics[angle=0,width=0.48\textwidth ]{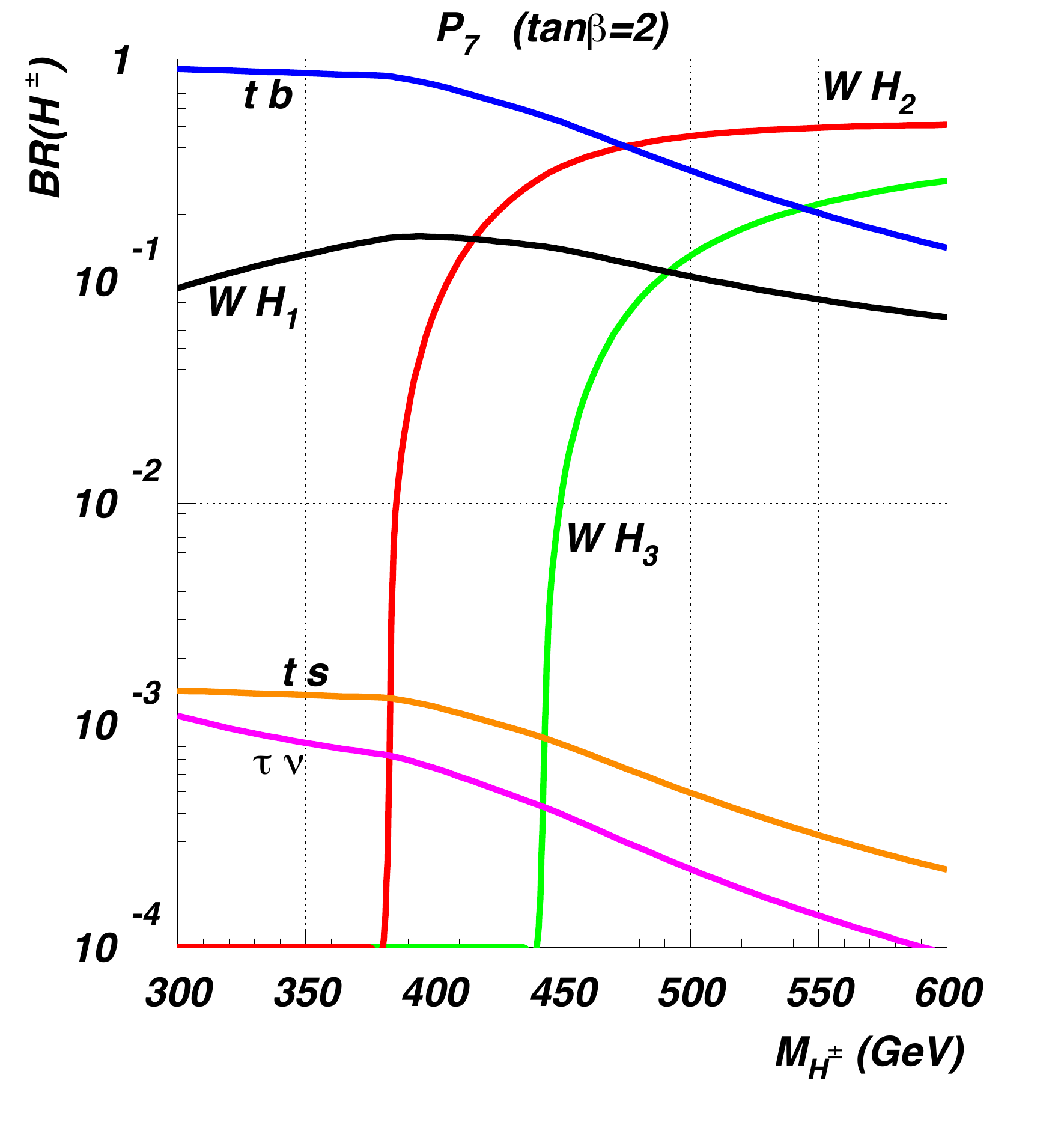}
  \includegraphics[angle=0,width=0.48\textwidth ]{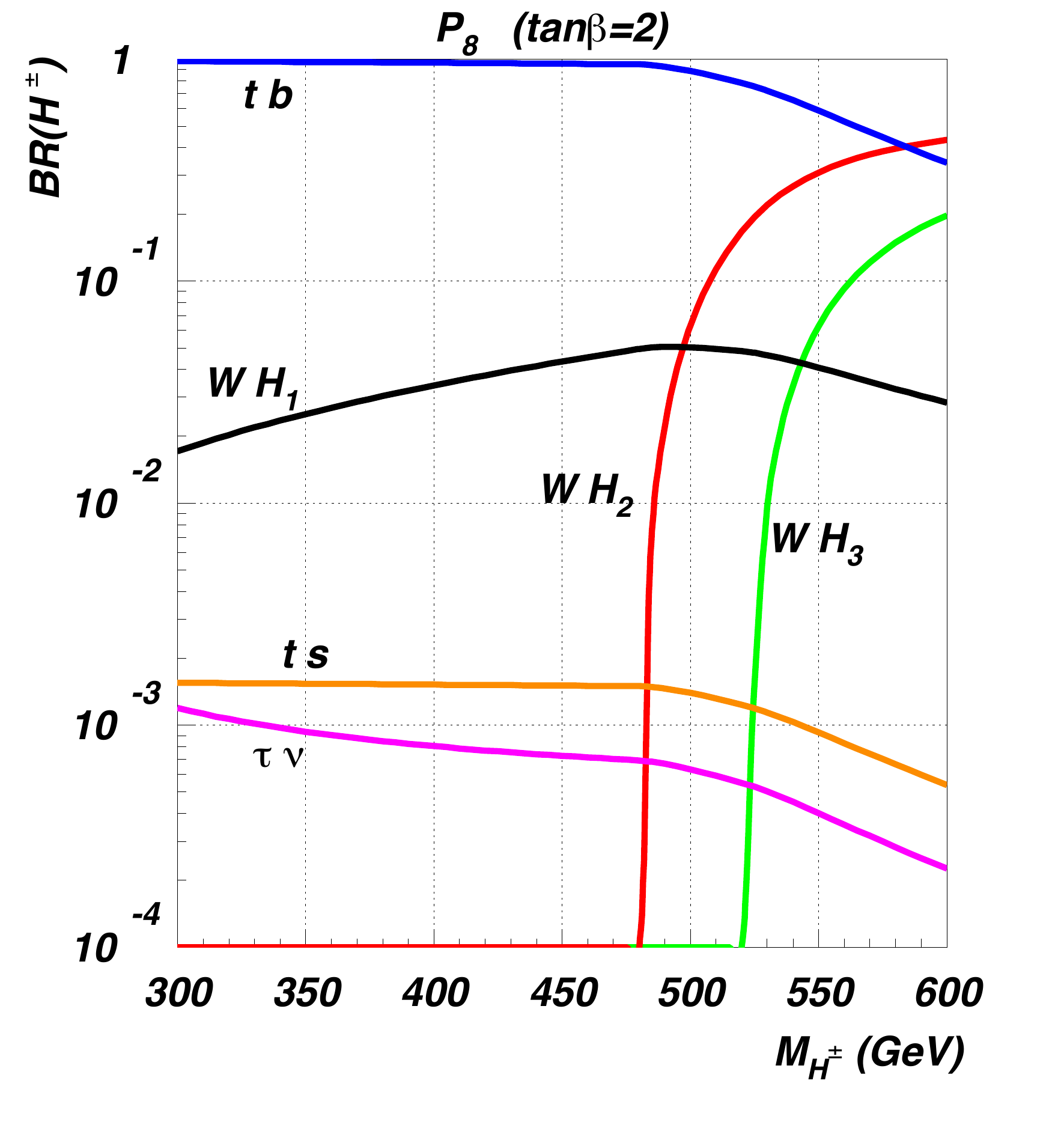}
\caption{Branching ratios of the charged Higgs vs mass. Four benchmark points have been considered, as indicated. 
\label{BR_Hc}}
\end{figure}

\subsection{Charged and lightest neutral Higgs bosons: BRs}

This subsection is devoted to an analysis of the BRs of the charged Higgs state in the allowed parts of the parameter space. In fact, since we are mainly interested in signatures of a charged Higgs boson produced in association with a $W$ boson, which involve model dependent couplings, it is of fundamental importance to establish some characteristic features of the BRs for some specific points of parameter space. In this regard, we consider four points from table~\ref{points} and we determine the most important decay modes. We consider only BRs~$>10^{-4}$, rates below this value are not of phenomenological relevance. Then, we have six decay modes: $W H_1$, $W H_2$, $W H_3$, $tb$, $ts$, $\tau \nu_\tau$, displayed in fig.~\ref{BR_Hc} for selected benchmark points. In addition, we remark that the results are presented for an illustrative range of $M_H=300-600$ GeV, while recalling that the allowed range is always bounded in the range $M_H\sim380-450$ GeV for an intermediate choice of $\tan\beta$.

First, we consider two points associated to the choice of $\tan\beta=1$ in fig.~\ref{BR_Hc}, $P_2$ and $P_6$. The dominant decay mode is always $tb$, and this feature is even reinforced when the masses $M_1$ and $M_2$ are closer ($P_2$), with respect to the case in which they are well separated ($P_6$). However, it is important to remark that the $W H_i$ branching fractions, when allowed by phase space, are always $\sim \mathcal{O}(0.1)$ and never smaller than $\sim \mathcal{O}(0.01)$. In particular, if $M_{H^\pm}>400$ GeV, then the BR for $W H_1$ is $\sim \mathcal{O}(0.1)$, this assures that the suppression brought about by this decay mode is never stronger than about an order of magnitude for a rather large value of $M_{H^\pm}$.

The result does not hold for the $W^\pm H_1$ case when $\tan\beta=2$ (see fig.~\ref{BR_Hc}, $P_7$ and $P_8$). In fact, it strongly depends on the choice of point in parameter space: for $P_8$ this decay mode is suppressed down to $\sim \mathcal{O}(0.01)$, while for $P_7$ its branching franction is restored to $\sim \mathcal{O}(0.1)$ because increasing the mixing between $\eta_1$ and $\eta_3$ via $|\alpha_2/\pi|=-0.03 \to -0.07$ increases its CP-odd component, and this effect leads to an enhancement.

Another feature of the $\tan\beta=2$ choice is that the $W^\pm H_2$ decay mode is always dominant as compared to the $tb$ one when $M_{H^\pm}$ is large (though not always allowed), due to the suppression of the $H^\pm \to tb$ coupling by a factor $\sim 2$ plus an always sizeable $H^\pm\to W^\pm H_2$ coupling. We remark that in this scenario the $W H_{1}$ BR is anyway $\sim \mathcal{O}(0.01)$ or bigger\footnote{The $\tau \nu_\tau$ decay mode, on the other hand, strongly depends on the $\tan\beta$ value. While it is not a primary aim of the present paper to analyse such a scenario, when $\tan\beta\sim \mathcal{O}(10)$ the $BR(H^\pm\to \tau \nu_\tau)$ could be enhanced up to $\sim \mathcal{O}(0.01)$.}.

Since we are interested in the phenomenology of the charged Higgs boson produced in association to vector bosons, it is important to understand the properties of the lightest neutral Higgs. For this, we conclude this subsection by presenting a relevant set of $H_1$ BRs. As is clear from table~\ref{BRH1}, the three most important decay mode are always the $b\bar{b}$, $gg$ and $WW^*$ ones. Since the first is the phenomenologically simplest among the three, we will only consider this decay channel for studying combined $H_1$ signatures.

\begin{table}[!t]
\begin{center}
\begin{tabular}{|c||c|c|c|c|c|}
\hline
 {\rm $H_1$ decay modes} & $P_2$ & $P_6$ & $P_7$ & $P_{8}$ \\ 
\hline 
$b\bar{b}$ & 0.3414 & 0.5916 & 0.2595 &  0.3697 \\ 
\hline 
$s\bar{s}$ & 0.0001 & 0.0002 & 0.0000 & 0.0001  \\ 
\hline 
$c\bar{c}$ & 0.0625 & 0.0317 & 0.0805 &0.0575  \\ 
\hline 
$\tau^+\tau^-$ & 0.0416 & 0.0721 & 0.0316 & 0.0451  \\ 
\hline 
$\mu^+\mu^-$ & 0.0001 & 0.0002 & 0.0001 & 0.0002  \\ 
\hline
$W^+W^-$ & 0.3158 & 0.1241 & 0.3218 & 0.3051  \\ 
\hline
$gg$ & 0.1944 & 0.1621 & 0.2617 & 0.1796  \\ 
\hline
$ZZ$ & 0.0386 & 0.0152 & 0.0393 & 0.0373  \\ 
\hline
$\gamma Z$ & 0.0024 & 0.0010 & 0.0024 &0.0023  \\ 
\hline
$\gamma \gamma$ & 0.0032 & 0.0018 & 0.0030 & 0.0031  \\ 
\hline
\end{tabular}
\end{center}
\caption{BRs of the lightest neutral Higgs, $\text{BR}(H_1\to X)$, for a set of benchmark points extracted from table~\ref{points}: $P_1$, $P_5$, $P_7$, $P_8$, $P_{9}$, $P_{10}$. \label{BRH1}}
\end{table}

\subsection{Charged Higgs: single production mechanisms}

\begin{figure}[!htb]
\subfloat[]{\label{ggHpWm}
\includegraphics[angle=0,width=0.7\textwidth ]{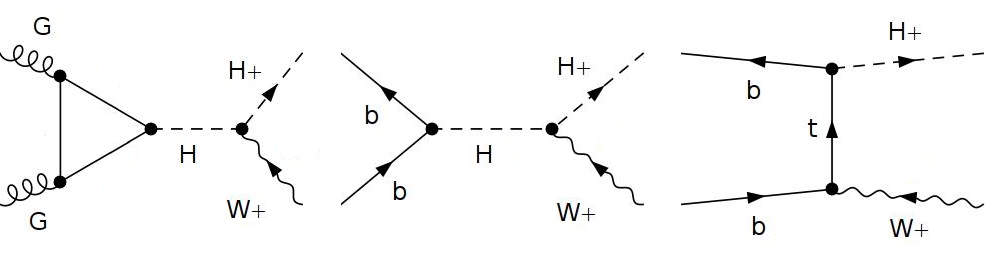}
} \\
\subfloat[]{\label{qqHpH}
\includegraphics[angle=0,width=0.21\textwidth ]{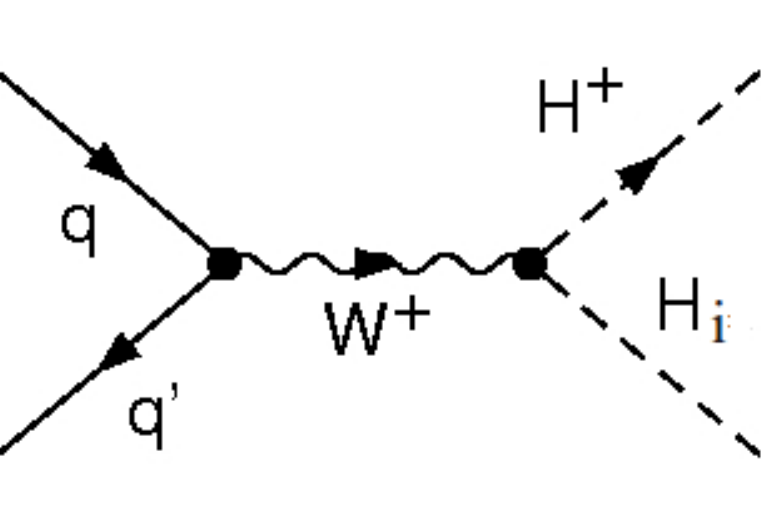}
} 
\qquad \qquad 
\subfloat[]{\label{ggHpbT}
\includegraphics[angle=0,width=0.45\textwidth ]{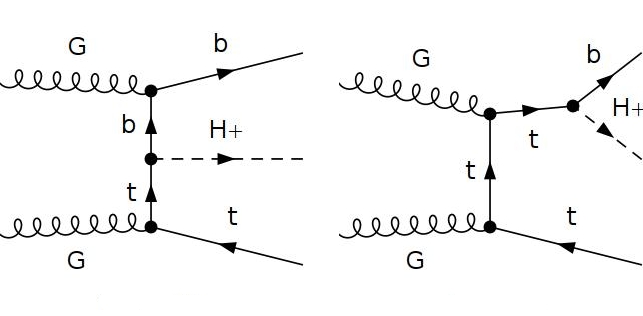}
}
\caption{Single charged Higgs production channels at parton level. \label{feynman}}
\end{figure}

In this subsection we study single charged Higgs production at the LHC for two choices of energy, $\sqrt{s}=8$ TeV and $\sqrt{s}=14$ TeV. Considering the partonic amplitudes, we have three main mechanisms to produce a single charged Higgs boson from hadrons, i.e., associated with bosons ({\bf B}) or fermions ({\bf F}): 
\begin{alignat}{3}
&{\bf B}: &\quad (gg,b\bar{b})&\rightarrow H^\pm W^\mp  &\quad &({\rm fig.~\ref{ggHpWm}}); \\
&{\bf B}: &\quad q\bar q'&\rightarrow H^\pm H_i  &\quad & ({\rm fig.~\ref{qqHpH}}); \\
&{\bf F}: &\quad gg&\rightarrow H^+b\bar{t} &\quad & ({\rm fig.~\ref{ggHpbT})}.
\end{alignat}
In fig.~\ref{feynman} we show the main partonic contributions to the three production channels.

Note that the process in fig.~\ref{qqHpH} will generally be disfavoured for two reasons. First, the quark density inside the proton is lower than the gluon density, so this channel is suppressed  in this respect, owing to the typical $H^\pm$ masses considered (recall that $x^2\propto M_{H^\pm}^2/s$, where $\sqrt{s}=8$ or 14 TeV). Second, the intermediate $W$ boson will be largely off-shell, also inducing significant depletion of the production rates. In contrast, the channel in fig.~\ref{ggHpWm} receives contributions from both quark and gluon initiated processes and can further be resonant in the $s$-channel (for some of our benchmarks). In principle, the box contribution from heavy fermions should be included in the set, for it has been proven in several scenarios \cite{Asakawa:2005nx,Dao:2010nu,Enberg:2011ae} that it can lead to $\mathcal{O}(10-100\%)$ corrections when $\tan\beta\sim \mathcal{O}(10)$. Still, in the phenomenological scenario that we propose to be tested at LHC ($\tan\beta\sim \mathcal{O}(1)$) this inclusion is totally irrelevant for our conclusions. The mode in fig.~\ref{ggHpbT} also benefits from counting on two subchannels, though it is never resonant (as $M_{H^\pm}>m_t-m_b$ for the model considered here).

\begin{figure}[!htb]
  \includegraphics[angle=0,width=0.48\textwidth ]{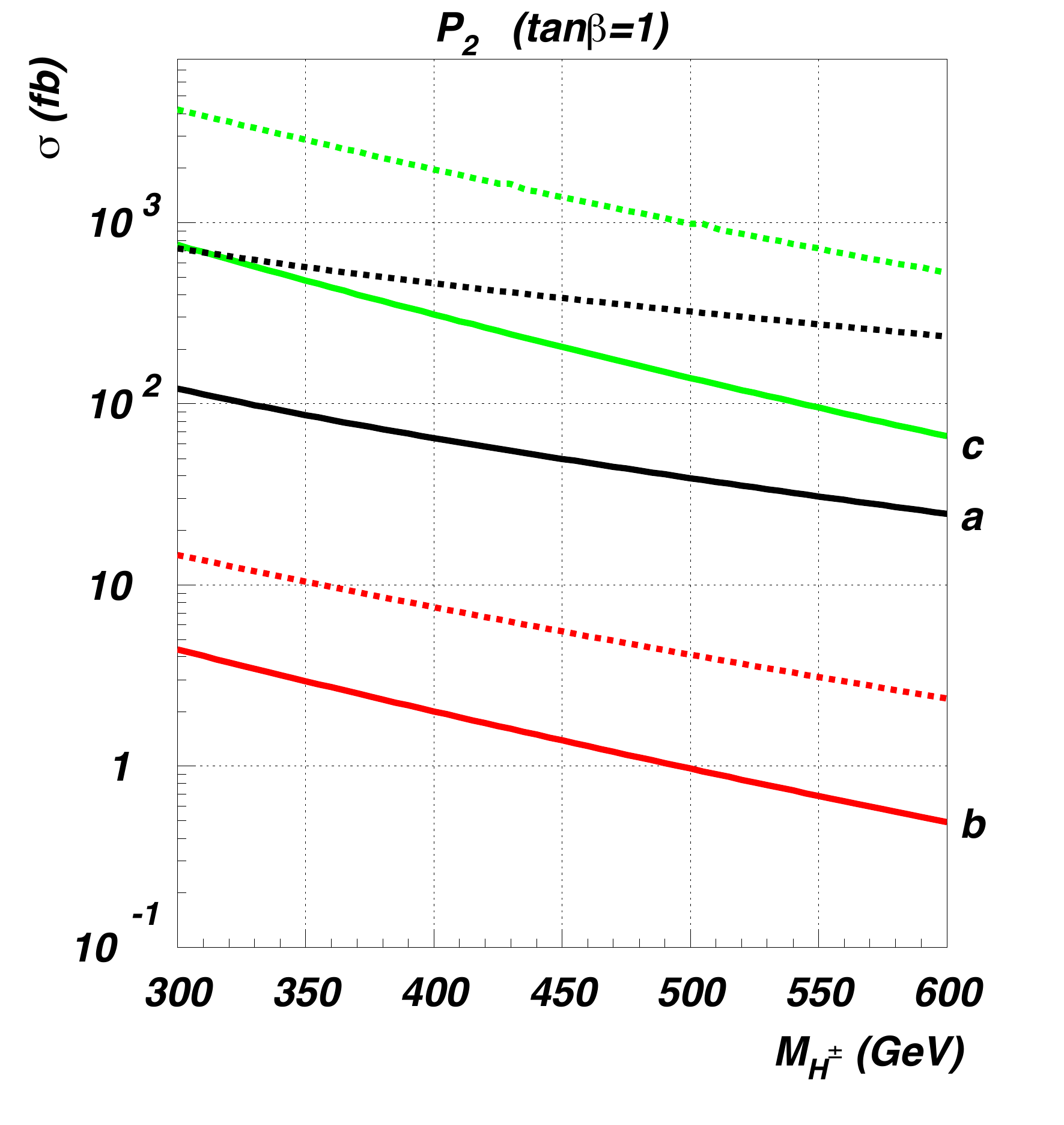}
  \includegraphics[angle=0,width=0.48\textwidth ]{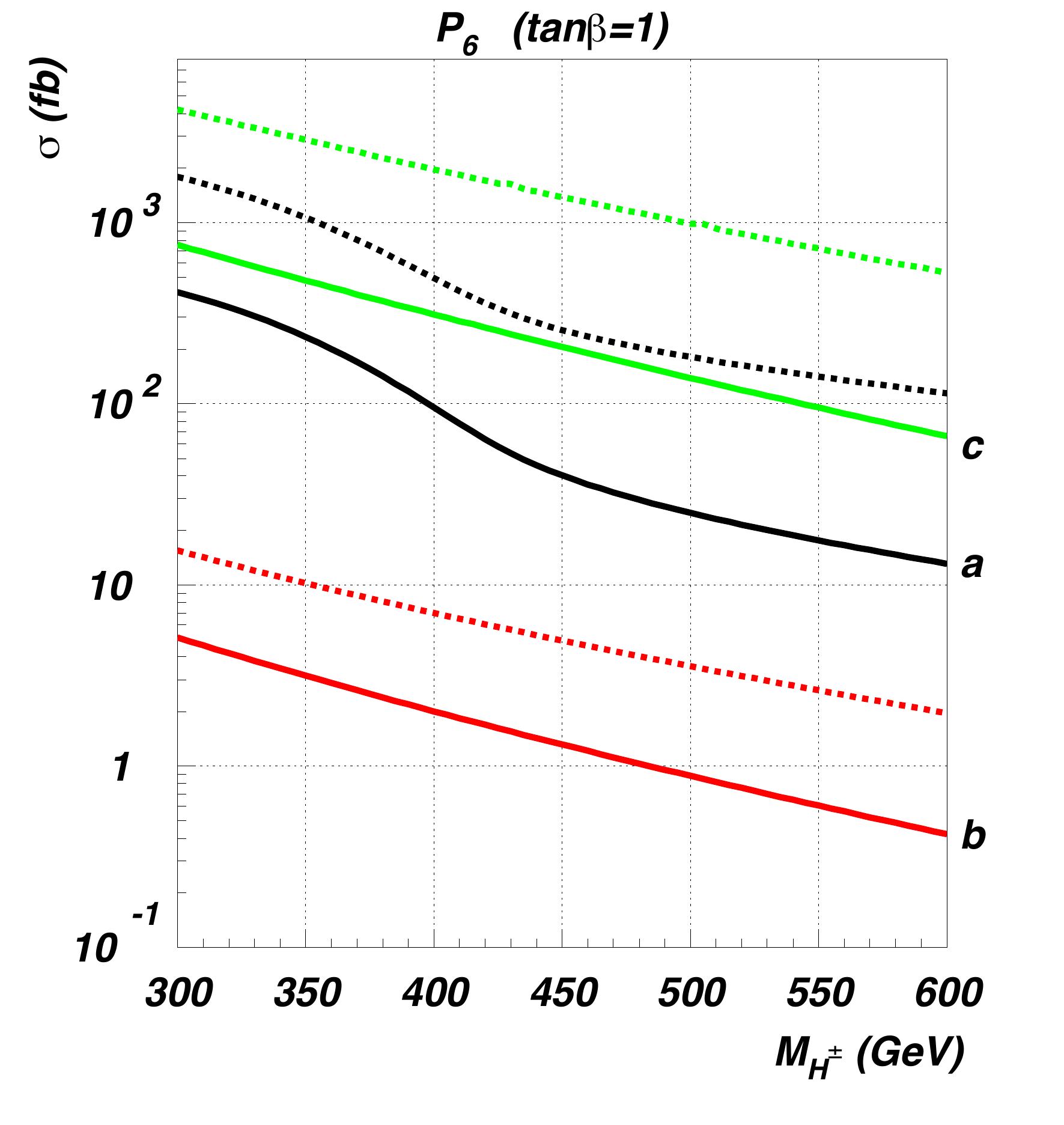}
  \includegraphics[angle=0,width=0.48\textwidth ]{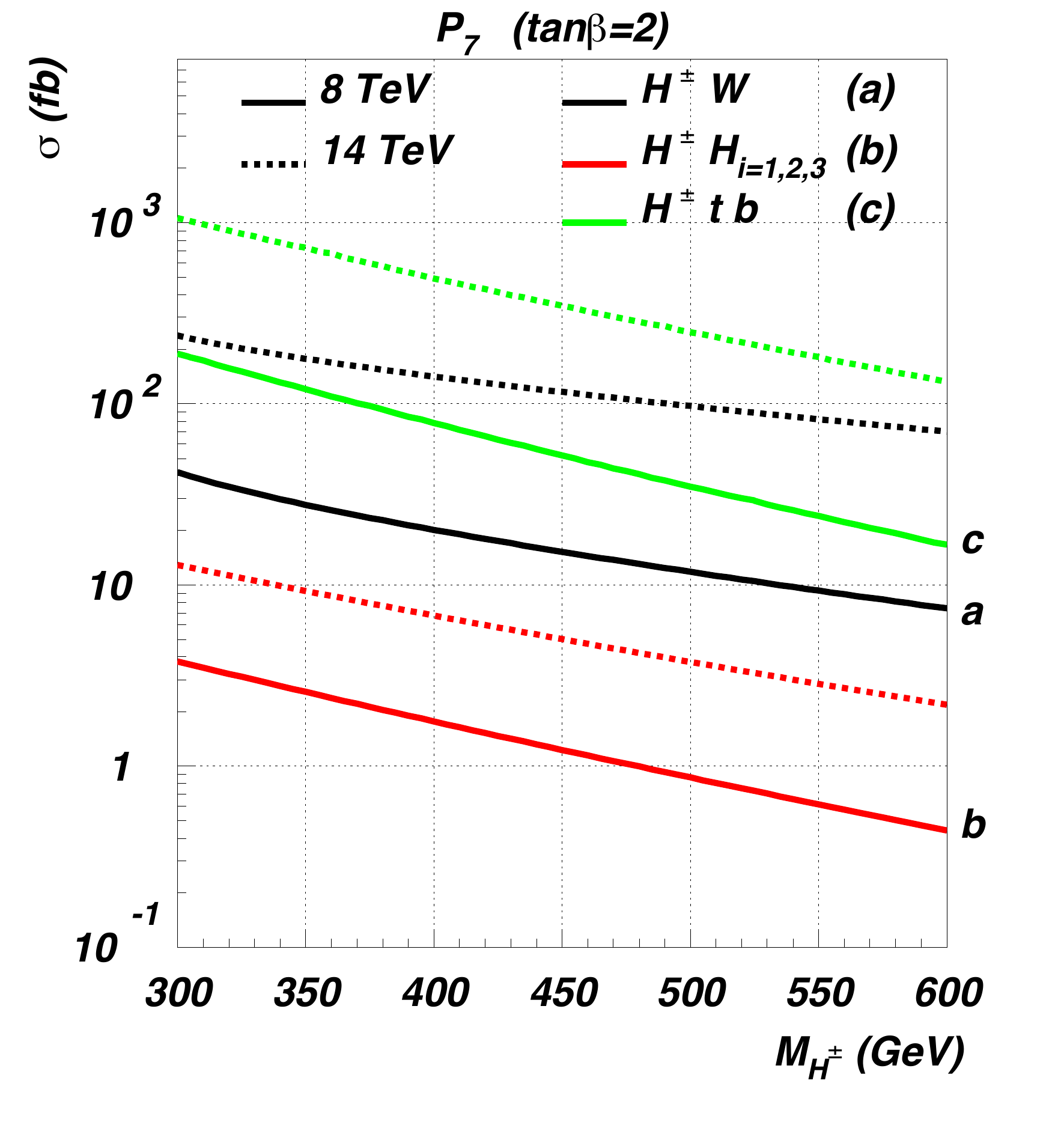}
  \includegraphics[angle=0,width=0.48\textwidth ]{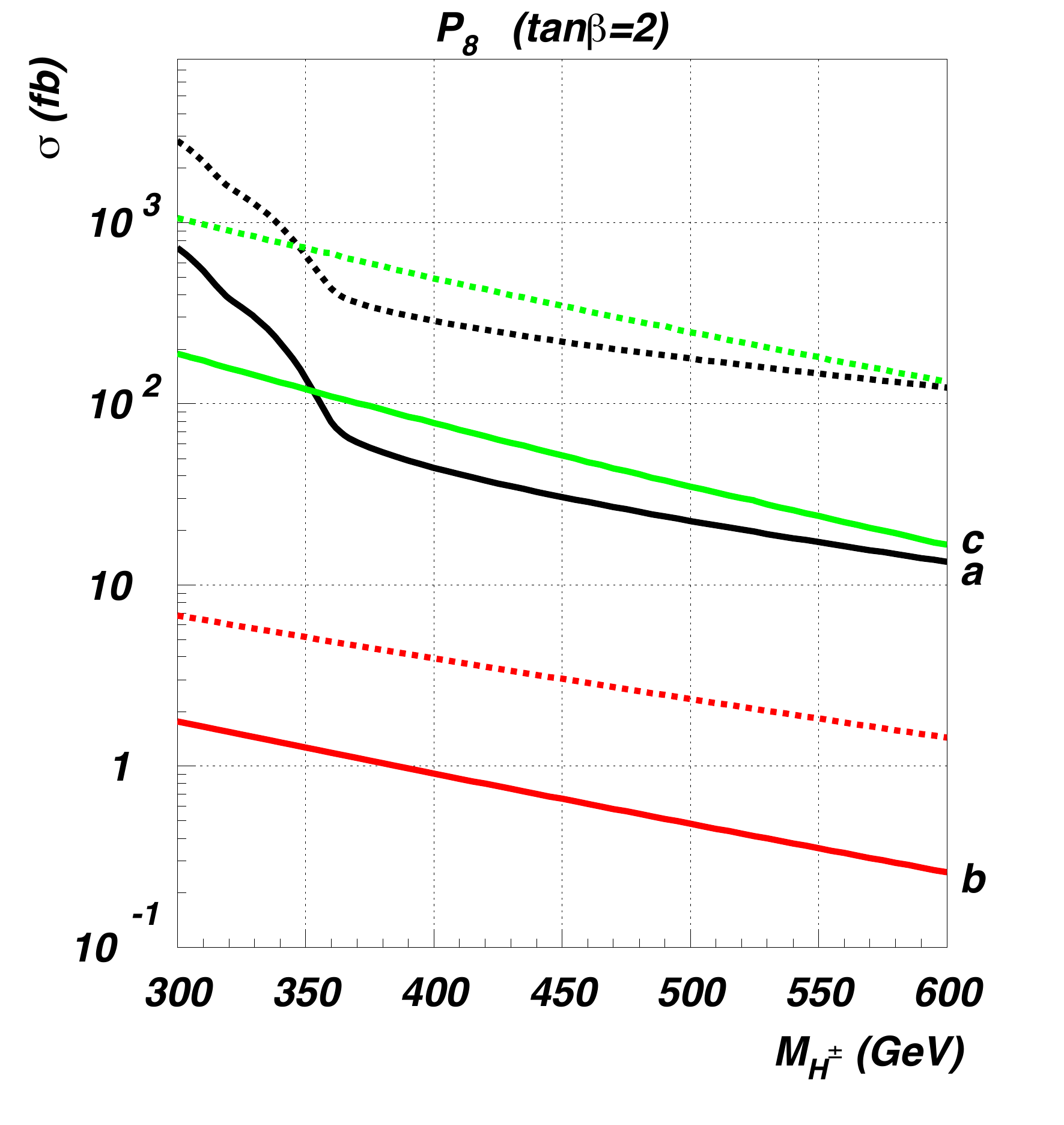}
\caption{Cross sections of the single production mechanisms vs the charged Higgs mass at $\sqrt{s}=8$ TeV (continuous lines) and $\sqrt{s}=14$ TeV (dashed lines). Four benchmark points are considered, as indicated. \label{sing_prod}}
\end{figure}

We consider now the four benchmark points $P_2$, $P_6$, $P_7$ and $P_8$ from table~\ref{points} and in fig.~\ref{sing_prod} we plot the cross sections against the charged Higgs mass for the aforementioned production channels. In the first place, we confirm that the associated production with a neutral scalar is disfavoured. Secondly, the remaining production mechanisms are always within a range of an order of magnitude at most. Again, we remark that the results are presented for an illustrative range of $M_H=300-600$ GeV, while recalling that the allowed range is always restricted to the range $M_H\sim380-450$ GeV for an intermediate choice of $\tan\beta$.

As regards the fermion-associated production mechanism of fig.~\ref{ggHpbT}, we remark that it only depends on the values of $M_{H^\pm}$ and $\tan\beta$ (see Eq.~(\ref{Eq:Yukawa-charged-II})), and there is a considerable reduction when moving from $\tan\beta=1$ to $\tan\beta=2$ (roughly a factor $2$) due to the fact that the dominant contribution in the coupling is $\sim m_t/\tan\beta$, hence the ratio of VEVs acts as a reduction factor. The cross section of the fermion-associated contribution at $\tan\beta=1$ is $\sim 10-10^2$ ($10^2-10^3$) fb when $\sqrt{s}=8$ ($14$) TeV and it is mostly inversely proportional to $\tan\beta$. 

The scope of the fermion-associated production mechanism in extracting a $H^\pm\to Wb\bar b$ signature (see below) has been analysed already in the literature, albeit in the MSSM, see \cite{Moretti:2000yg}, and we will revisit it in a CP-violating type-II 2HDM in a future publication. 

Instead, here, we concentrate on vector-boson-associated production. The corresponding cross sections show a complicated behaviour with respect to different choices of parameters. We start our analysis by considering the channel with a final $H^\pm W^\mp$ state. From fig.~\ref{sing_prod} ($P_2$ and $P_6$) we see that a choice of $\tan\beta=1$ plus a low-Higgs-masses scenario ($P_2$: $M_2 \sim M_3\sim 300-400$ GeV) has a cross section $\sim 10-10^2$ ($10^2-10^3$) fb when $\sqrt{s}=8$ ($14$) TeV, and that it is dominant (competitive) with respect to the fermion-associated production. On the other hand, we see that the choice of a high $M_2$ ($P_6$) and $M_3$ ($\geq 500$ GeV) favour the contribution from the parton-level channel $gg\to H_{i=2,3} \to H^\pm W^\mp$ proceeding through the on-shell $H_{i=2,3}$, and this results in a cross section that is always dominant and even enhanced when $m_W+M_{H^{\pm}}<M_{i=2,3}$, i.e. $\sim 10^2-10^3$ ($10^3-10^4$) fb when $\sqrt{s}=8$ ($14$)~TeV.

These qualitative conclusions also hold when $\tan\beta=2$. Despite an overall suppression by one order of magnitude due to the increased value of $\tan\beta$, from fig.~\ref{sing_prod} ($P_7$) we see that on-shell production is not taking place (the $H_2$ and $H_3$ masses are too light). Hence, starting from a cross section of $\sim 10^2$ ($10^3$) fb when $\sqrt{s}=8$ ($14$) TeV at the lower $M_{H^\pm}$ scale we find a rate of $\sim 10$ ($10^2$) fb when $\sqrt{s}=8$ ($14$) TeV for high values of $M_{H^\pm}$. However, the on-shell production is important for $P_{8}$, where we can see an interplay between the on-shell $H_2$ and $H_3$ ($M_3\sim 460$ GeV) production at $M_{H^\pm}\sim 300$ GeV being realised by a ``double-shoulder'' shaped line. In fact, when $M_{H^\pm}\sim 320$ ($380$) GeV the $H_2$ ($H_3$) on-shell production is switched off. In this framework, the cross section is $\sim 10$ ($10^2$) fb when $\sqrt{s}=8$ ($14$) TeV, but it is increased by an order of magnitude when $M_{H^\pm}< 380$ GeV.

The high-$\tan\beta$ benchmark points give rather low production cross sections. For $P_9$ and $P_{10}$ we find cross sections roughly half and a tenth, respectively, of that for $P_8$. Since the latter, after the cuts discussed below, does not yield any useful signal, we have not explored $P_9$ and $P_{10}$ any further.

Finally, we briefly comment on the sub-dominant channel with a $W$-mediated $H^\pm H_{i=1,3}$ final state. This channel is unlikely to have interesting phenomenological implication at the LHC, at least at the early running stage: when $\sqrt{s}=8$ TeV and with integrated luminosity $\Lumint=10~\text{fb}^{-1}$ the cross section is typically just above the threshold for producing a few events. Since it is not competitive with the other production channels, we will not study this mechanism.

In the next subsection we consider the neutral-Higgs-mediated production mechanism for analysing the scope of the LHC in discovering such a state in the allowed parameter space. 

\subsection{$pp\rightarrow H^\pm W^\mp$: significance analysis}

In this subsection we analyse the significance of single charged Higgs boson production in association with gauge bosons for the set of benchmark points in table~\ref{points} (except $P_9$ and $P_{10}$). All figures herein refer to an integrated luminosity of $100$ fb$^{-1}$.

Among the different charged Higgs decay modes, we have chosen to study $H^\pm \to W^\pm H_i$. The decay chain with the $H_i\to b\bar b$ intermediate decay is numerically favoured, so that we adopt it here, hence the complete $H^\pm$ decay chain is 
\begin{equation}
H^\pm \to W^\pm H_1 \to W^\pm b\bar{b}.
\end{equation}

Therefore, we are interested in the significance ($\Sigma\sim S/\sqrt{B}$) analysis of a $2b+2W$ final state produced via a single charged Higgs. The most important  background at the LHC for this final state is top quark pair production. However, we will show that a systematic reduction of this background is possible.

Notice that one may worry here about the contribution of the fermionic charged Higgs decay chain
\begin{equation}
H^+  \to  t \bar{b} \to  W^+ b \bar{b},
\end{equation}
as it yields an irreducible final state that is identical to the chosen one that could be defined as part of either the signal or the background. Under any circumstances, though, we believe that our top-mass veto (see below) will render this contribution negligible, so we omit it here\footnote{We will attempt extracting this particular $H^\pm$ topology in the context of the CP-violating 2HDM in a separate publication, with the aim of improving upon the MSSM results obtained in \cite{Moretti:1998xq}.}.

A $b$-tagging efficiency of $\sim 70\%$ has been assumed for each $b$-(anti)quark in the final state, and a full reconstruction efficiency has been assumed with respect to the $W$ bosons. Among the possible decay patterns of the two $W$ bosons, the semileptonic one was chosen, i.e., one hadronic and one leptonic decay, allowing for the full reconstruction of the events (unlike the fully leptonic decay mode) and a neater environment than the fully hadronic decay
mode.

Hence, the overall selected process for the signal is the following: 
\begin{equation}
pp\to W^\mp H^\pm \to W^\mp W^\pm H_1 \to W^\mp W^\pm b \bar{b} \to 2j+2b+1\ell + \mbox{MET}.
\end{equation}

For each benchmark point, $2 \cdot 10^4$ unweighted events were produced. Regarding the top background, $4.5 \cdot 10^6$ unweighted events (with generation cuts) have been simulated in CalcHEP. For both signal and background the standard set of CTEQ6.6M \cite{Nadolsky:2008zw} PDFs with scale $Q=\sqrt{s}$ were employed. For emulating a real LHC-prototype detector, a Gaussian smearing was included to take into account the electromagnetic energy resolution of $0.15/\sqrt{E}$ and the hadronic energy resolution of $0.5/\sqrt{E}$.

We describe now the overall strategy for the background reduction procedure. A first set of  cuts includes typical detector kinematic acceptances and standard intermediate object reconstruction, such as $W\to jj$ and $H_1 \to b\overline{b}$ (cuts 1--3). Further, a $t$-(anti)quark reconstruction is used as ``top veto'' (cut~4). Led by the consideration that a $b$ quark pair stemming from the Higgs boson is boosted (unlike the almost back-to-back pair from $t\overline{t}$), we define the last cut of the following set (cut~5):

\begin{itemize}
\item[1) ] {\bf Kinematics:} standard detector cuts
\begin{alignat}{2}
p^T_\ell &> 15 \mbox{ GeV}, &\qquad \left|\eta _\ell\right| &< 2.5, \nonumber\\
p^T_j& > 20 \mbox{ GeV}, &\qquad \left|\eta _j\right| &< 3, \\
\left| \Delta R_{jj} \right|&> 0.5, &\qquad \left| \Delta R_{\ell j} \right|&> 0.5; \nonumber
\end{alignat}
with $\eta$ the pseudorapidity and $\Delta R=\sqrt{(\Delta\eta)^2+(\Delta\phi)^2}$.
\item[2) ] {\bf light Higgs reconstruction:}
\begin{equation}
\left| M(b\overline{b}) - 125  \mbox{ GeV}\right|< 20  \mbox{ GeV} \, ;
\end{equation}
\item[3) ] {\bf hadronic $W$ reconstruction ($W_h\to jj$):}
\begin{equation}
\left| M(jj) - 80  \mbox{ GeV}\right|< 20  \mbox{ GeV} \, ;
\end{equation}
\item[4) ] {\bf top veto}:
if $\Delta R (b_1, W_h) < \Delta R (b_2, W_h)$, then
\begin{equation}
 M(b_1 jj) > 200  \mbox{ GeV} \, , \qquad  M_T(b_2 \ell\nu) > 200  \mbox{ GeV}\, ,
\end{equation}
otherwise $1\leftrightarrow 2$; 
\item[5) ] {\bf same-hemisphere $b$ quarks:}
\begin{equation}
 \frac{{\bf p}_{b_1}}{|{\bf p}_{b_1}|} \cdot  \frac{{\bf p}_{b_2}}{|{\bf p}_{b_2}|}  > 0 \, .
\end{equation}
\end{itemize}

In table~\ref{cuts} we show the efficiency of the previous set of cuts against the simulated background for the $P_2$ and $P_4$ points of table~\ref{points}, for two $H^\pm$ masses. There is a clear correlation between the $M_{H^\pm}$ value and the efficiency of the top veto (the most effective cut of this set): the higher the mass, the higher the efficiency.

\begin{table}[!ht]
\begin{center}
\begin{tabular}{|l|c||c|c||c|c||c|}
\hline
\multicolumn{2}{|c|}{ Cut } & BG events & BG Eff. ($\%$) & $P_2$ events & $P_2$ Eff. ($\%$) & $S/\sqrt{B}$\\ 
\hline
1: &Kin. & $310291$ &$100$ & $\;\;54.2, \,59.4\;\, $ & $100, \,100 $ & $0.1, \,0.1 $\\
\hline
2:&$H_1$ rec.  & $263629$ &$85.0$ & $53.8, \,59.0$ & $99.3, \,99.3$ & $0.1, \,0.1$  \\
\hline
3:&$W$ rec. & $256745$ &$97.4$ & $52.2, \,57.4$ & $97.0, \,97.3$ & $0.1, \,0.1$ \\
\hline
4:&top veto & $1689$ &$0.7$ & $15.5, \,29.3$ & $29.7, \,51.0$ & $0.4, \,0.7$  \\
\hline
5:&same-side $b$'s & $708$ &$41.9$ & $11.8, \,22.8$ & $76.1, \,77.9$ & $0.4, \,0.9$  \\
\hline
\end{tabular}

\begin{tabular}{|l|c||c|c||c|c||c|}
\hline
\multicolumn{2}{|c|}{ Cut }  & BG events & BG Eff. ($\%$) & $P_4$ events & $P_4$ Eff. ($\%$) & $S/\sqrt{B}$\\ 
\hline
1: &Kin. & $310291$ &$100$ & $356.2, \,166.4$ & $100, \,100$ & $0.6, \,0.3$\\
\hline
2: &$H_1$ rec. & $263629$ &$85.0$ & $351.5, \,165.7$ & $98.7, \,99.6$ & $0.7, \,0.3$  \\
\hline
3: &$W$ rec. & $256745$ &$97.4$ & $341.9, \,160.4$ & $97.3, \,96.8$ & $0.7, \,0.3$ \\
\hline
4: &top veto & $1689$ &$0.7$ & $41.6, \,70.3$ & $12.2, \,43.8$ & $1.0, \,1.7$  \\
\hline
5: &same-side $b$'s & $708$ &$41.9$ & $32.8, \,54.7$ & $78.7, \,77.7$ & $1.2, \,2.1$  \\
\hline
\end{tabular}
\end{center}
\caption{Consecutive efficiencies of the cuts imposed on the top quark background and (top) on the $P_2$ point, and (bottom) on the $P_4$ point (with $M_{H^\pm}=310, \,390$ GeV) of table~\ref{points}. \label{cuts}}
\end{table}

\begin{figure}[!htb]
  \includegraphics[angle=0,width=0.66\textwidth ]{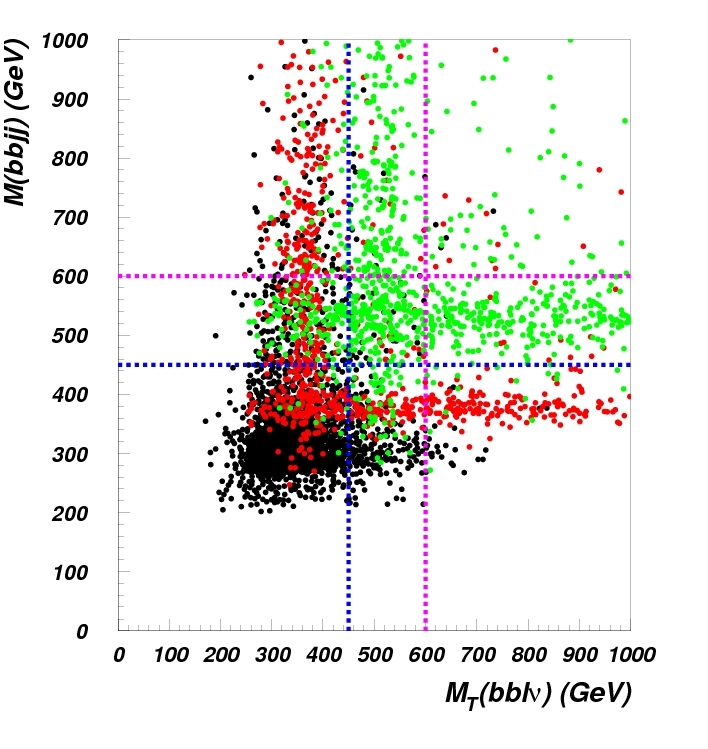}
\caption{$M(b\overline{b}jj)$ vs. $M_T(b\overline{b}\ell\nu)$ after cut~5 for (unweighted) point $P_2$, with $M_{H^\pm}=375$ GeV (red, bottom-left) and $M_{H^\pm}=525$~GeV (green, top-right). In black is the (unweighted) top background. The dashed lines show $M_{\rm{lim}}= 450$ and 600~GeV. \label{2D_cut}}
\end{figure}

After these rather generic cuts are imposed, more signal-based selections can improve the significance.
The main consideration of the following analysis is that the charged Higgs mass can equivalently be reconstructed by either the invariant mass of the four jets ($2b+2j$), $M(b\overline{b}jj)$, or the transverse mass of the $b$ jets, the lepton and the MET, $M_T(b\overline{b}\ell\nu)$. Let's focus on the $M(b\overline{b}jj)$--$M_T(b\overline{b}\ell\nu)$ plane: for the signal, either of the two variables (if not both) will always reconstruct the correct charged Higgs boson mass, thus producing a cross-like shape in the plane defined by the
two masses. In contrast, the background events accumulate at $\sim 2m_t$, as can be seen in fig.~\ref{2D_cut} in which we adopt an illustrative choice of charged Higgs masses.

The presence of long tails for the signal towards regions where the top background is heavily reduced allows us to introduce two specific (and alternative) cuts: 
\begin{eqnarray}\label{cut_sq}
\mbox{``squared cut'': }\qquad \mbox{C}_\text{squ} &=& \max\big( M(b\overline{b}jj),M_T(b\overline{b}\ell \nu) \big)>M_{\rm{lim}} \, \\ \label{cut_sg}
\mbox{``single cut'': }\qquad \mbox{C}_\text{sng} &=& M_T(b\overline{b}\ell\nu)>M_{\rm{lim}}\, .
\end{eqnarray}
The single cut of eq.~(\ref{cut_sg}) is applied only on $M_T(b\overline{b}\ell\nu)$ because the reduction of the top background is higher than if compared to a similar cut on the $M(b\overline{b}jj)$ for the same numerical value of $M_{\rm{lim}}$. 

To determine which is the better of the two proposed strategies and what is the optimal value for $M_{\rm{lim}}$, we studied the effects of C$_\text{squ}$ and C$_\text{sng}$ for several values of $M_{\rm{lim}}$. Results are shown in tables~\ref{cuts_p2} and \ref{cuts_p4} for the points $P_2$ and $P_4$, respectively.

\begin{table}[ht]
\begin{center}
\begin{tabular}{|c|c||c||c|c||c|c|}
\hline
\multicolumn{2}{|c||}{\multirow{2}{*}{Cut}} & $t\overline{t}$ & \multicolumn{2}{|c||}{$P_2=310$ GeV} & \multicolumn{2}{|c|}{$P_2=390$ GeV}  \\ \cline{3-7}
\multicolumn{2}{|c||}{} & Events & Events & $S/ \sqrt{B}$ & Events & $S/\sqrt{B}$  \\ 

\hline
\multirow{2}{*}{$M_{\rm{lim}}=450$ GeV} &
    C$_\text{sng}$ & $66.6$  & $6.6$  & $0.8$ & $12.2$ & $1.5$\\

\ & C$_\text{squ}$ & $161.1$ & $10.5$ & $0.8$ & $20.1$ & $1.6$\\
\hline
\multirow{2}{*}{$M_{\rm{lim}}=500$ GeV} &
    C$_\text{sng}$ & $45.2$  & $6.0$ & $0.9$ & $11.1$ & $1.6$ \\

\ & C$_\text{squ}$ & $118.8$ & $9.7$ & $0.9$ & $18.4$ & $1.7$ \\
\hline
\multirow{2}{*}{$M_{\rm{lim}}=550$ GeV} &
    C$_\text{sng}$ & $30.3$ & $5.1$ & $0.9$ & $9.9$  & $1.8$ \\

\ & C$_\text{squ}$ & $91.0$ & $8.5$ & $0.9$ & $16.1$ & $1.7$ \\
\hline
\multirow{2}{*}{$M_{\rm{lim}}=600$ GeV} &
    C$_\text{sng}$ & $24.9$ & $4.7$ & $1.0$ & $8.9$ & $1.8$ \\

\ & C$_\text{squ}$ & $63.1$ & $7.7$ & $1.0$ & $14.3$ & $1.8$ \\
\hline
\end{tabular}
\end{center}
\caption{Comparison between C$_\text{squ}$ and C$_\text{sng}$ vs $M_{\rm{lim}}$ for $P_2$: surviving events and significance with respect to the background. \label{cuts_p2}}
\end{table}

\begin{table}[ht]
\begin{center}
\begin{tabular}{|c|c||c||c|c||c|c|}
\hline
\multicolumn{2}{|c||}{\multirow{2}{*}{Cut}} & $t\overline{t}$ & \multicolumn{2}{|c||}{$P_4=310$ GeV} & \multicolumn{2}{|c|}{$P_4=390$ GeV}  \\ \cline{3-7}
\multicolumn{2}{|c||}{} & Events & Events & $S/ \sqrt{B}$ & Events & $S/\sqrt{B}$ \\ 

\hline
\multirow{2}{*}{$M_{\rm{lim}}=450$ GeV} &
    C$_\text{sng}$ & $66.6$  & $14.5$ & $1.8$ & $29.0$ & $3.6$\\

\ & C$_\text{squ}$ & $161.1$ & $25.8$ & $2.0$ & $47.3$ & $3.7$\\
\hline
\multirow{2}{*}{$M_{\rm{lim}}=500$ GeV} &
    C$_\text{sng}$ & $45.2$  & $12.7$ & $1.9$ & $26.3$ & $3.9$\\

\ & C$_\text{squ}$ & $118.8$ & $22.4$ & $2.1$ & $43.0$ & $3.9$\\
\hline
\multirow{2}{*}{$M_{\rm{lim}}=550$ GeV} &
    C$_\text{sng}$ & $30.3$ & $10.8$ & $2.0$ & $23.4$ & $4.2$\\

\ & C$_\text{squ}$ & $91.0$ & $19.8$ & $2.1$ & $37.9$ & $4.0$\\
\hline
\multirow{2}{*}{$M_{\rm{lim}}=600$ GeV} &
    C$_\text{sng}$ & $24.9$  & $10.0$ & $2.0$ & $20.3$ & $4.1$\\

\ & C$_\text{squ}$ & $63.1$  & $17.7$ & $2.2$ & $33.1$ & $4.2$\\
\hline
\end{tabular}
\end{center}
\caption{Comparison between C$_\text{sng}$ and C$_\text{squ}$ vs $M_{\rm{lim}}$ for $P_4$: surviving events and significance with respect to the background. \label{cuts_p4}}
\end{table}

Clearly, a higher value for $M_{\rm{lim}}$ results in an increase of the significance, the top background is reduced more than the signal. It is important to note that for low charged Higgs masses, C$_\text{squ}$ seems to perform better than the single cut. However, this is strickly true for $M_{H^\pm}\simeq 310$ GeV only: if a further selection is imposed, restricting the evaluation of the significance to the peak-region only
\begin{equation}\label{cut_peak}
\mbox{peak cut:}\qquad \left|M-M_{H^\pm}\right|<50\mbox{ GeV}\, ,
\end{equation}
the significance obtained by imposing C$_\text{sng}$, when calculated for all the other charged Higgs boson mass values, is \emph{always higher} than the one obtained by imposing C$_\text{squ}$. Here, $M=\min\big( M(b\overline{b}jj),M_T(b\overline{b}\ell \nu) \big)$ when eq.~(\ref{cut_sq}) is employed, while $M=M(b\overline{b}jj)$ when eq.~(\ref{cut_sg}) is employed.

For the following analysis, the value $M_{\rm{lim}}=600$ GeV has been chosen as well as the selection C$_\text{sng}$, this choice provides the best significance and a narrower peak while keeping a sufficient number of signal events ($>10$). Should the surviving signal events be less than $10$, it would then be advisable to choose instead the squared cut C$_\text{squ}$ for the higher survival probability of the signal events (despite the lower significance and the broader peak).

\begin{figure}[!htb]
  \includegraphics[angle=0,width=0.475\textwidth ]{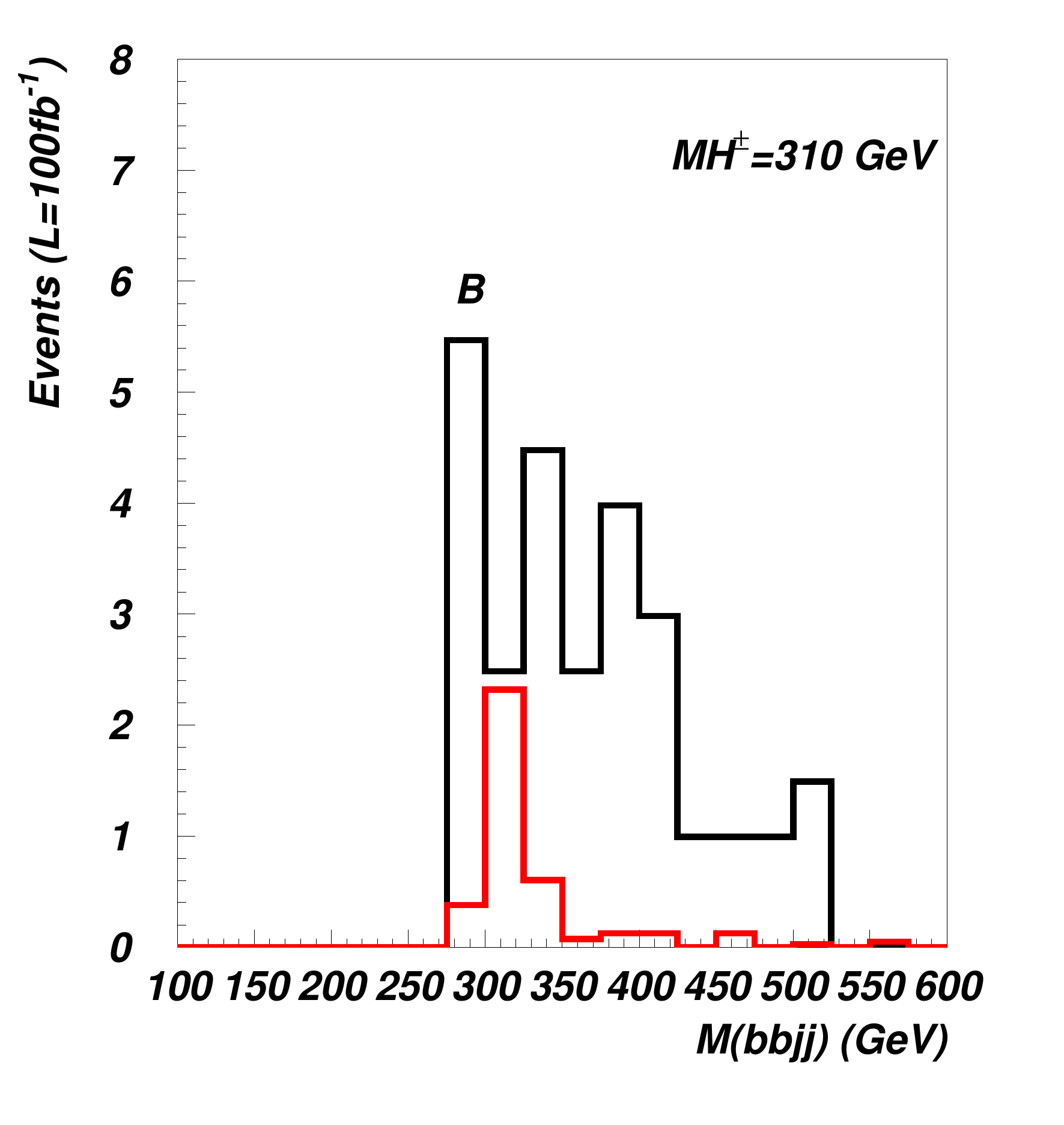}
  \includegraphics[angle=0,width=0.475\textwidth ]{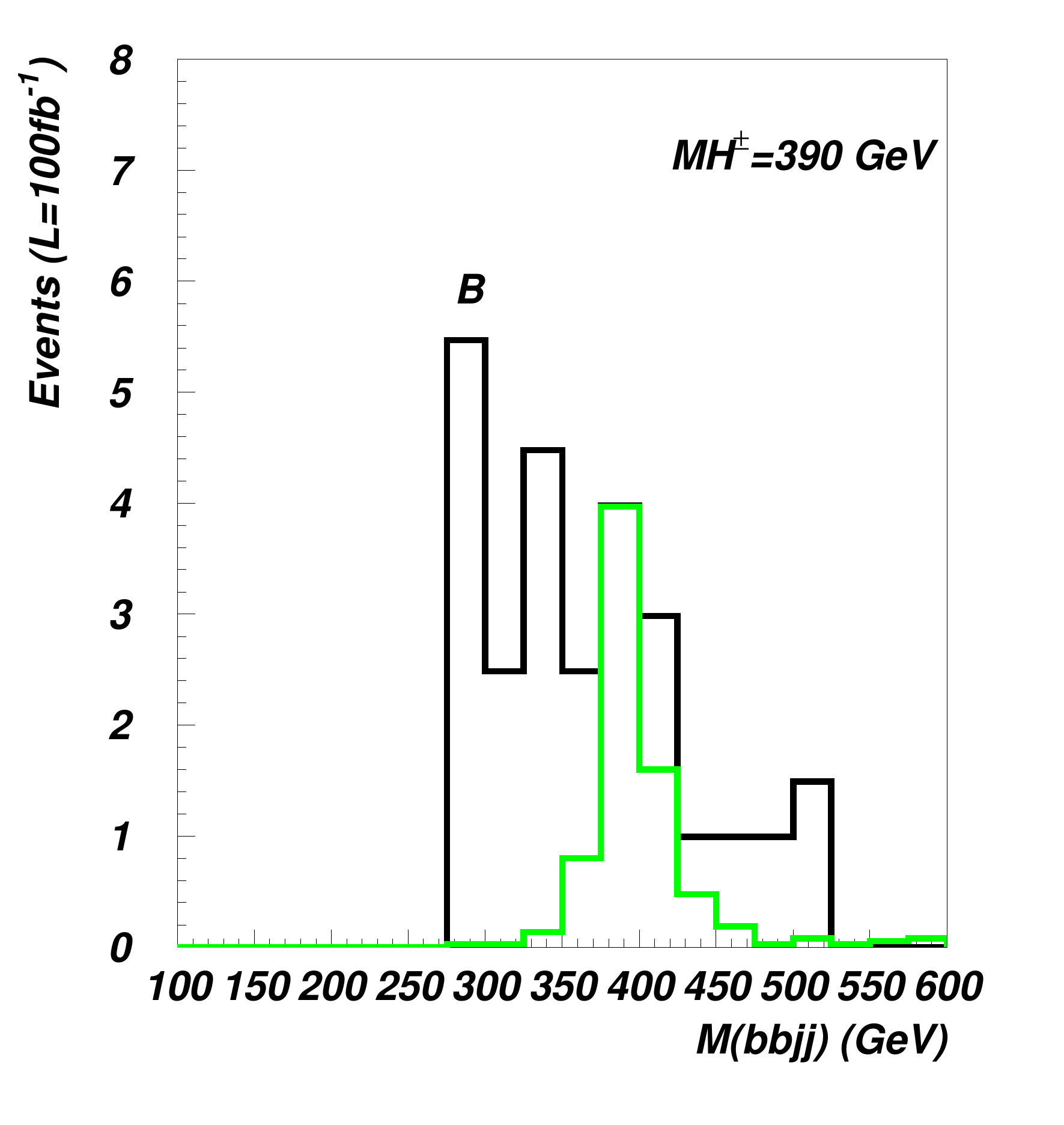}
\caption{Point $P_2$. Number of events integrated with $\Lumint=100$ fb$^{-1}$ at $\sqrt{s}=14$ TeV vs $M(b\overline{b}jj)$ for signal (coloured lines) and $t$-quark background (black histogram, labeled ``B''). \label{Pi_2}}
\end{figure}

\begin{figure}[!htb]
  \includegraphics[angle=0,width=0.475\textwidth ]{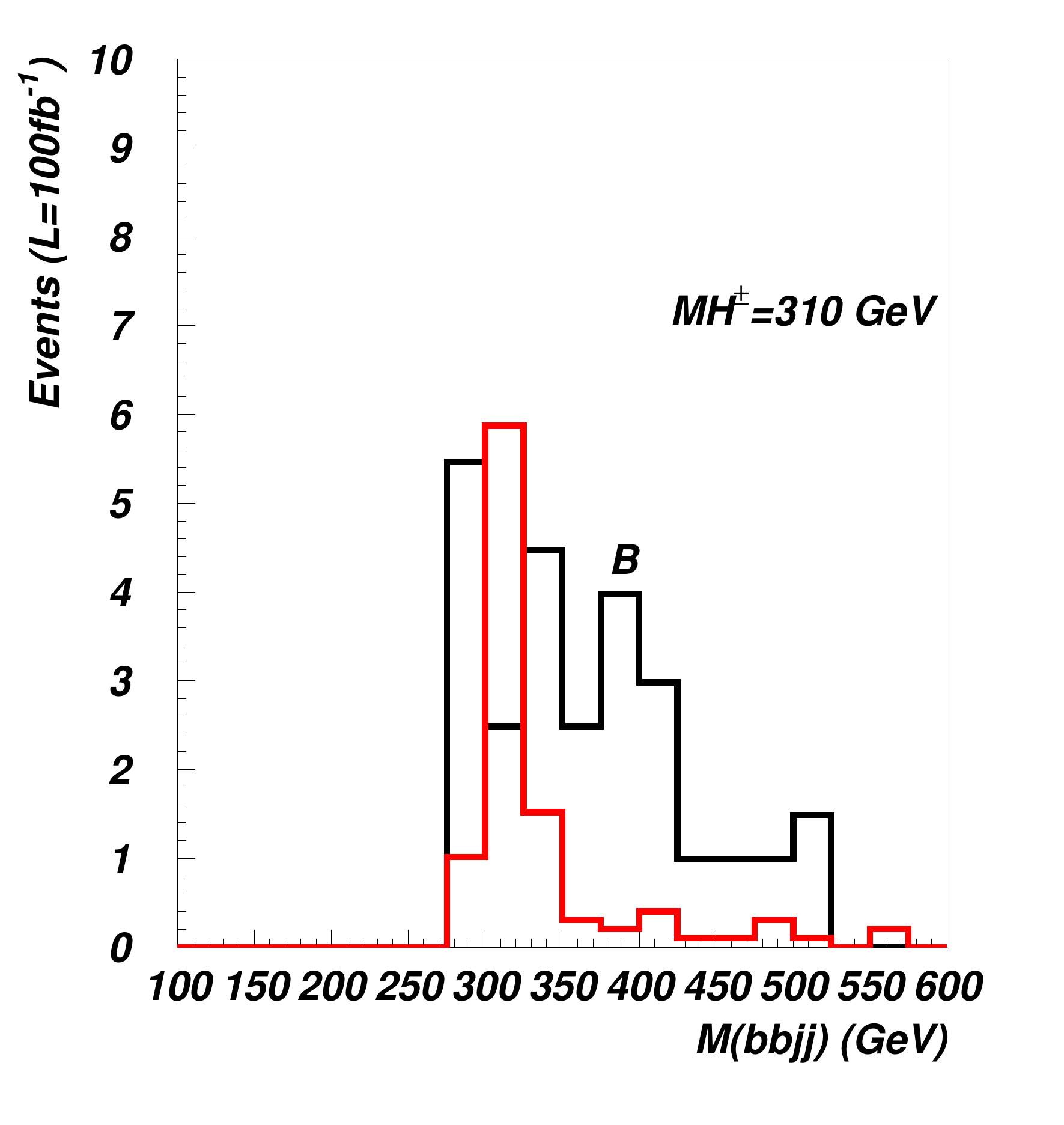}
  \includegraphics[angle=0,width=0.475\textwidth ]{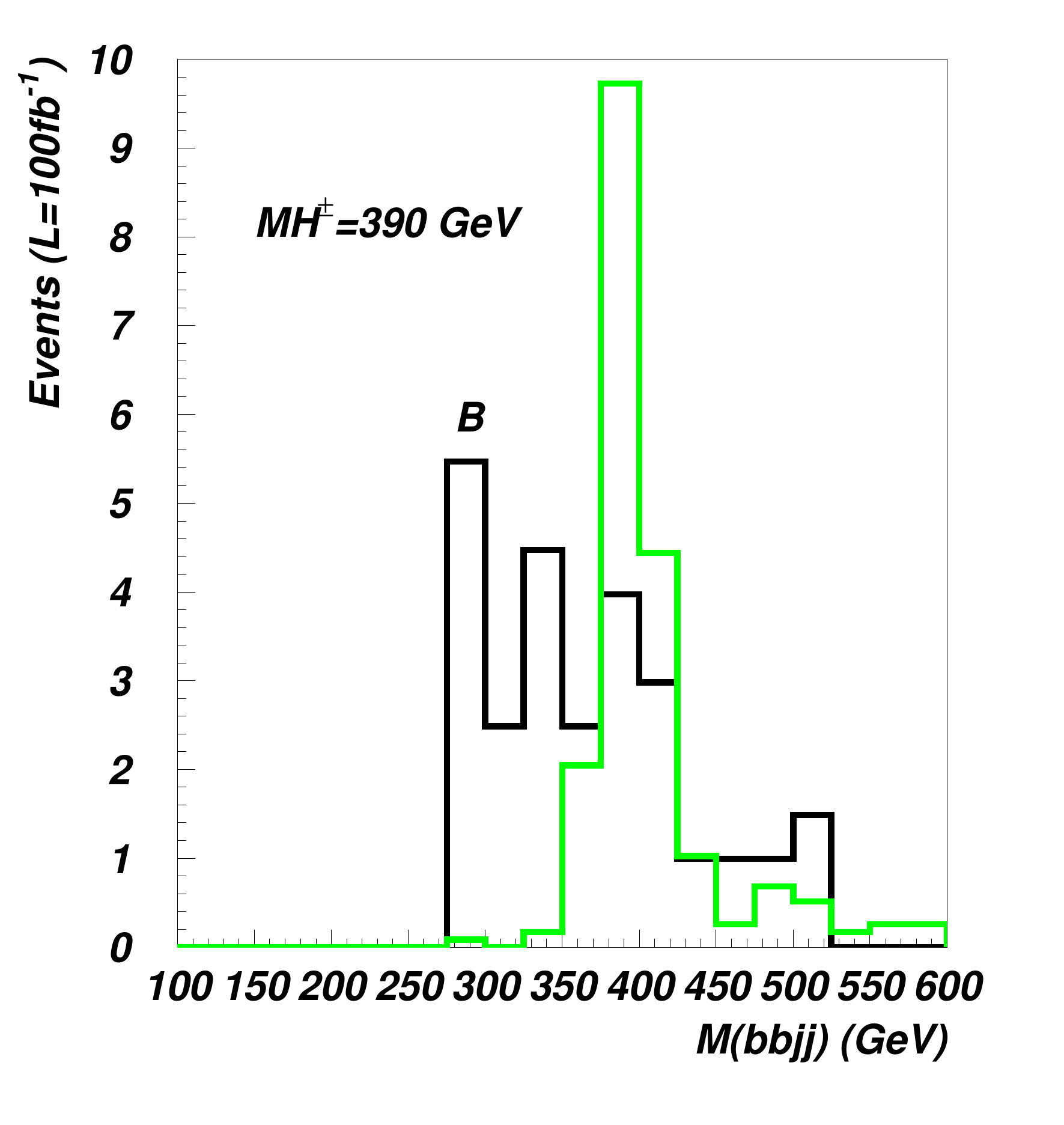}  
\caption{Point $P_3$. Similar to fig.~\ref{Pi_2}. \label{Pi_3}}
\end{figure}

\begin{figure}[!htb]
  \includegraphics[angle=0,width=0.475\textwidth ]{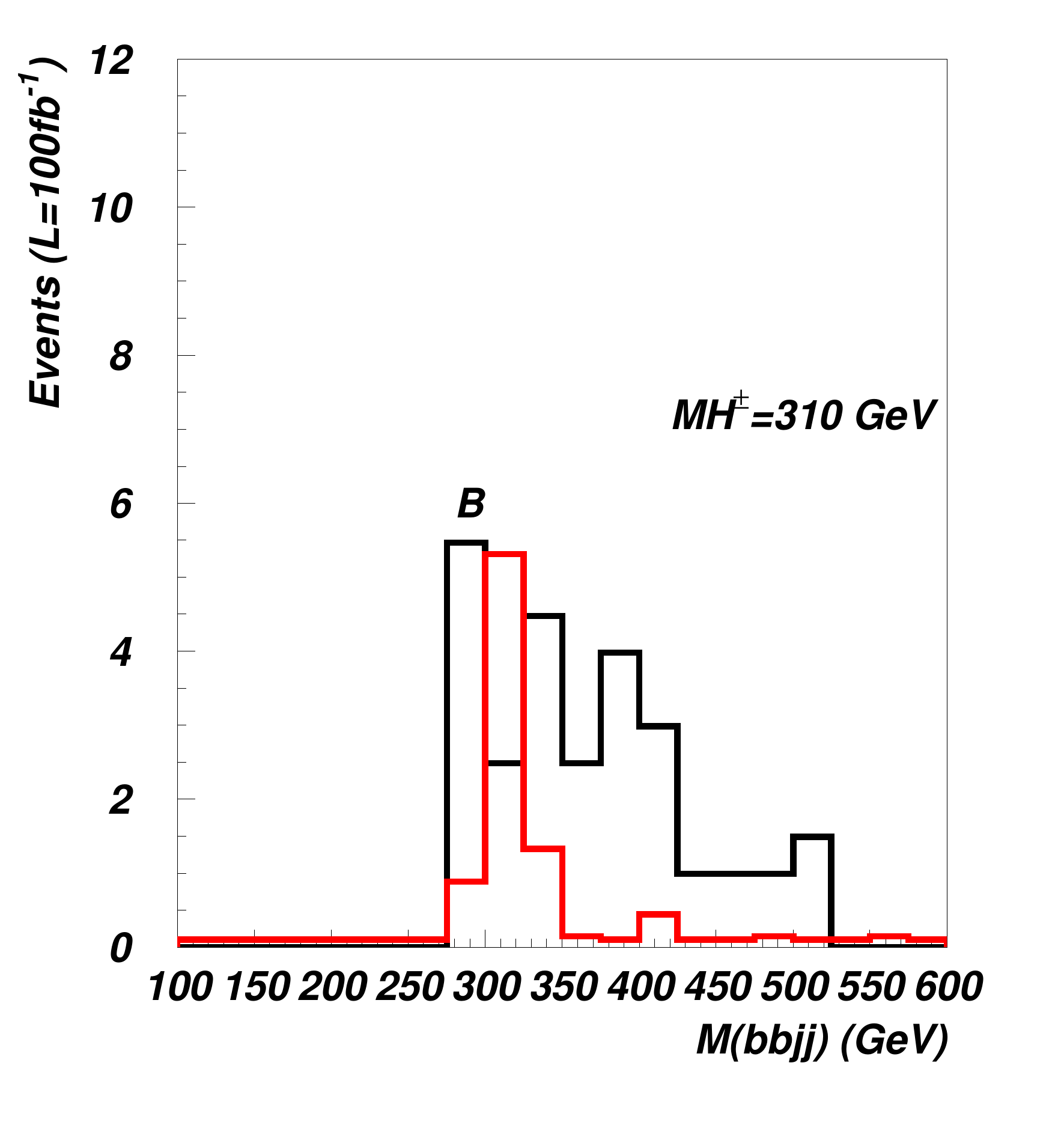}
  \includegraphics[angle=0,width=0.475\textwidth ]{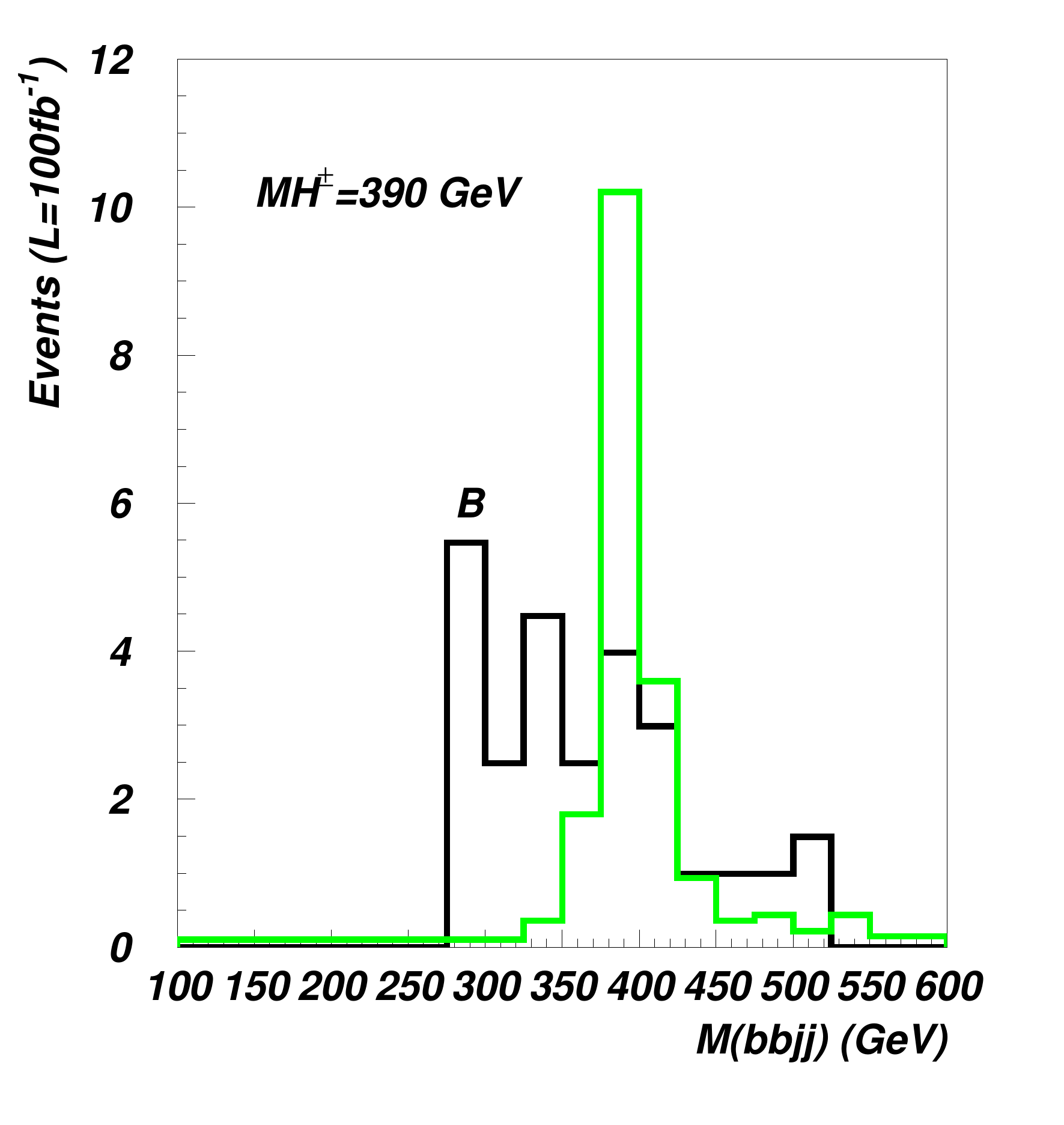}  
\caption{Point $P_4$. Similar to fig.~\ref{Pi_2}. \label{Pi_4}}
\end{figure}

\begin{figure}[!htb]
  \includegraphics[angle=0,width=0.475\textwidth ]{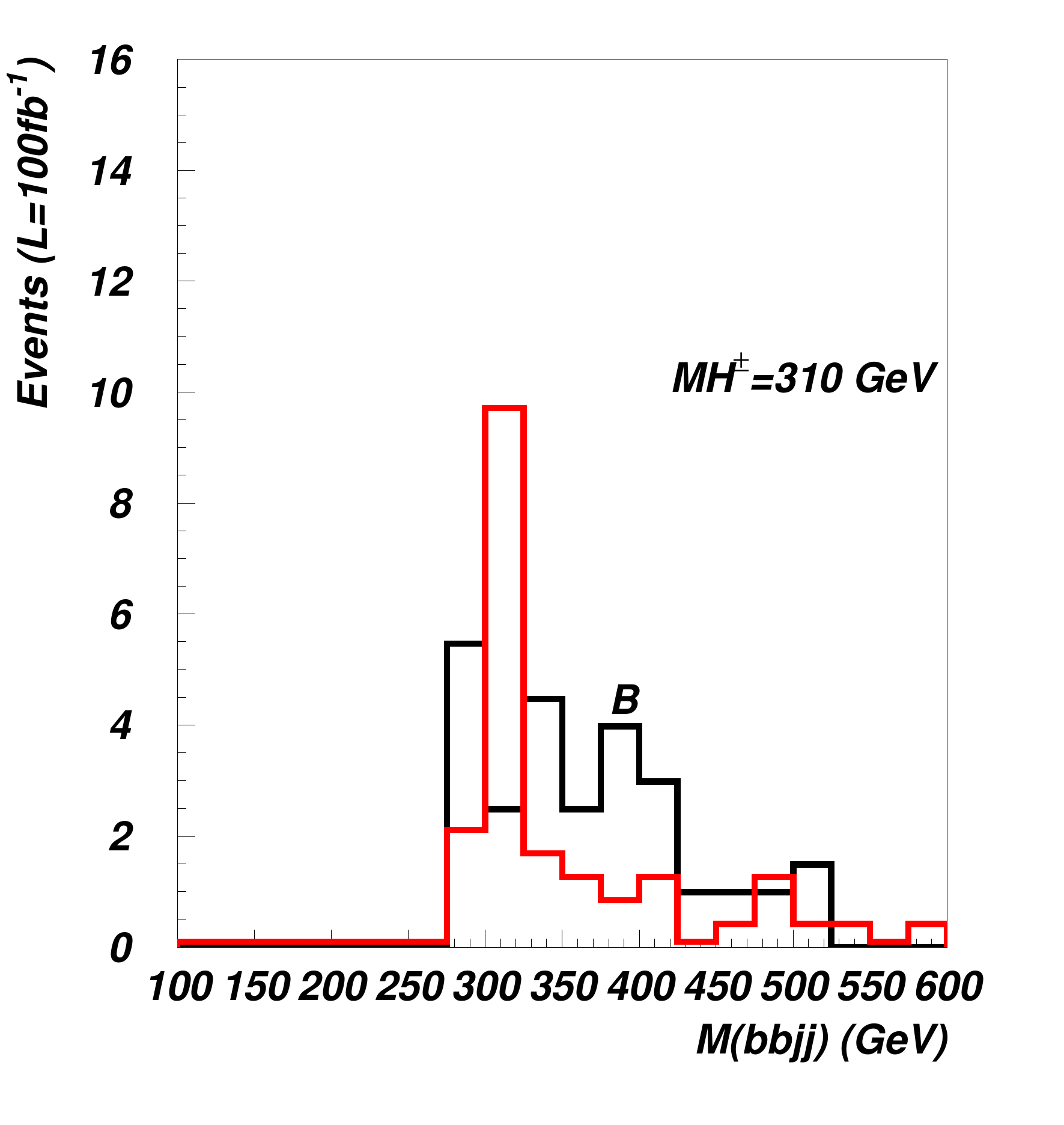}
  \includegraphics[angle=0,width=0.475\textwidth ]{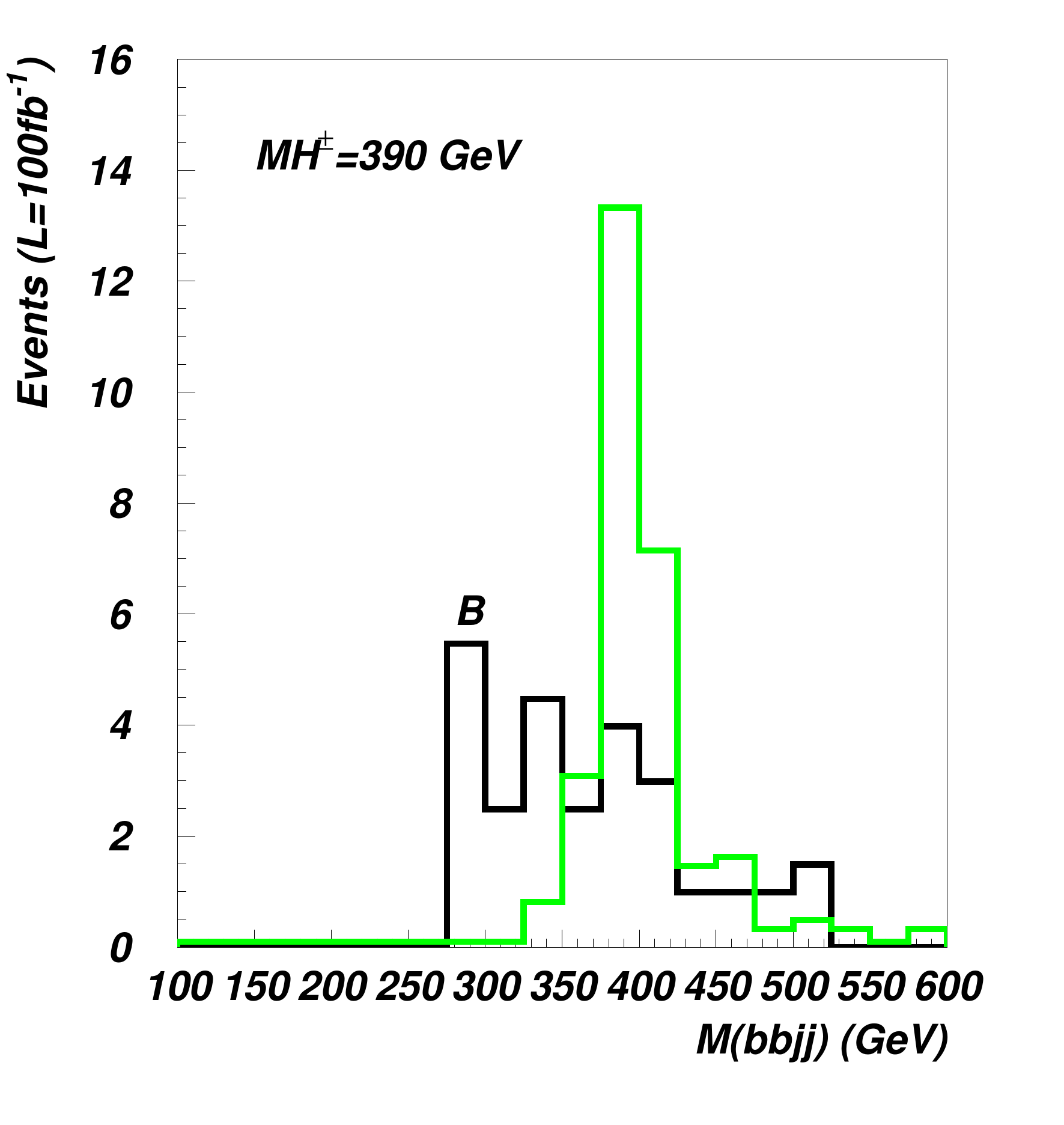}  
\caption{Point $P_5$. Similar to fig.~\ref{Pi_2}. \label{Pi_5}}
\end{figure}

\begin{figure}[!htb]
  \includegraphics[angle=0,width=0.475\textwidth ]{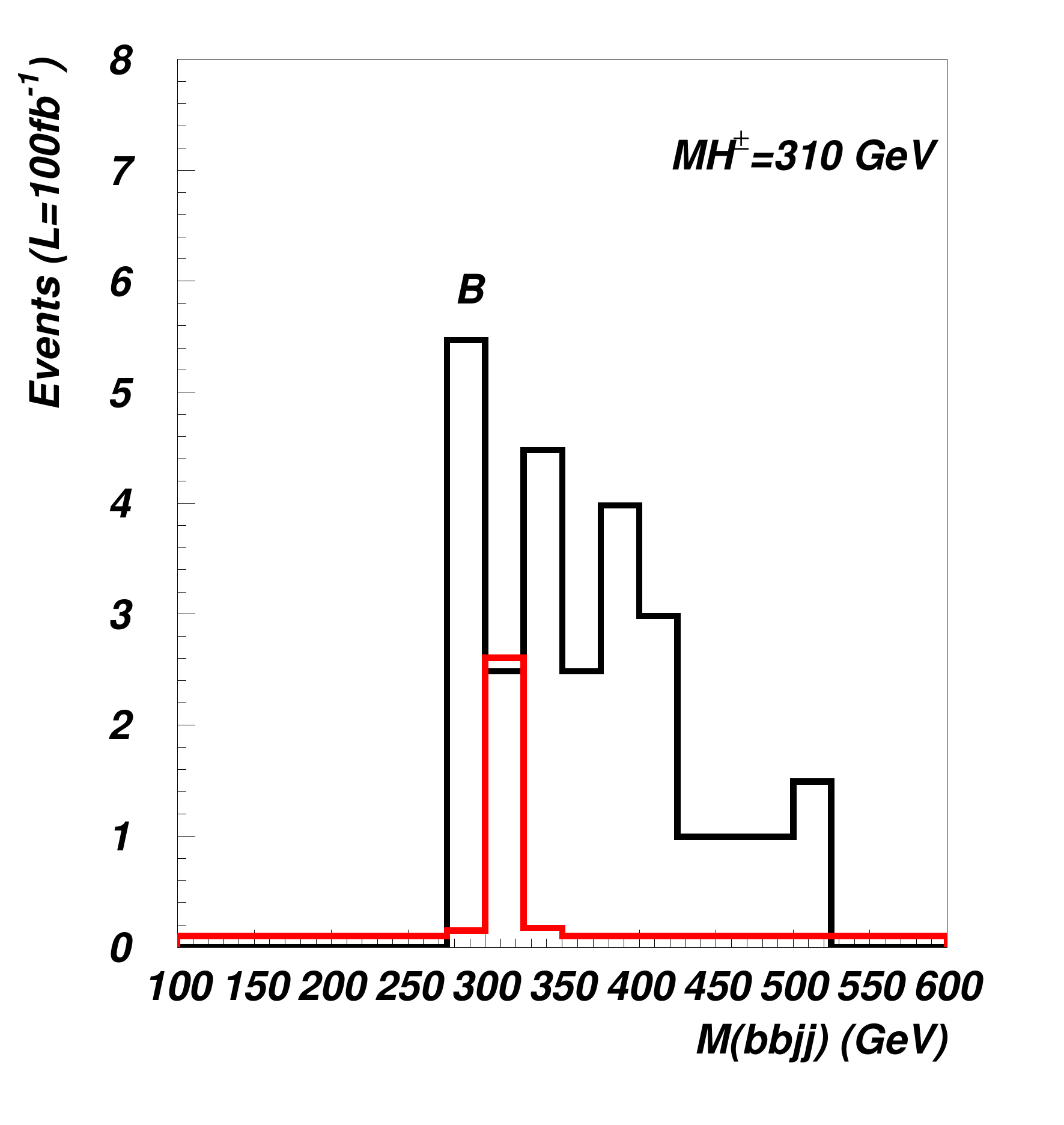}  
  \includegraphics[angle=0,width=0.475\textwidth ]{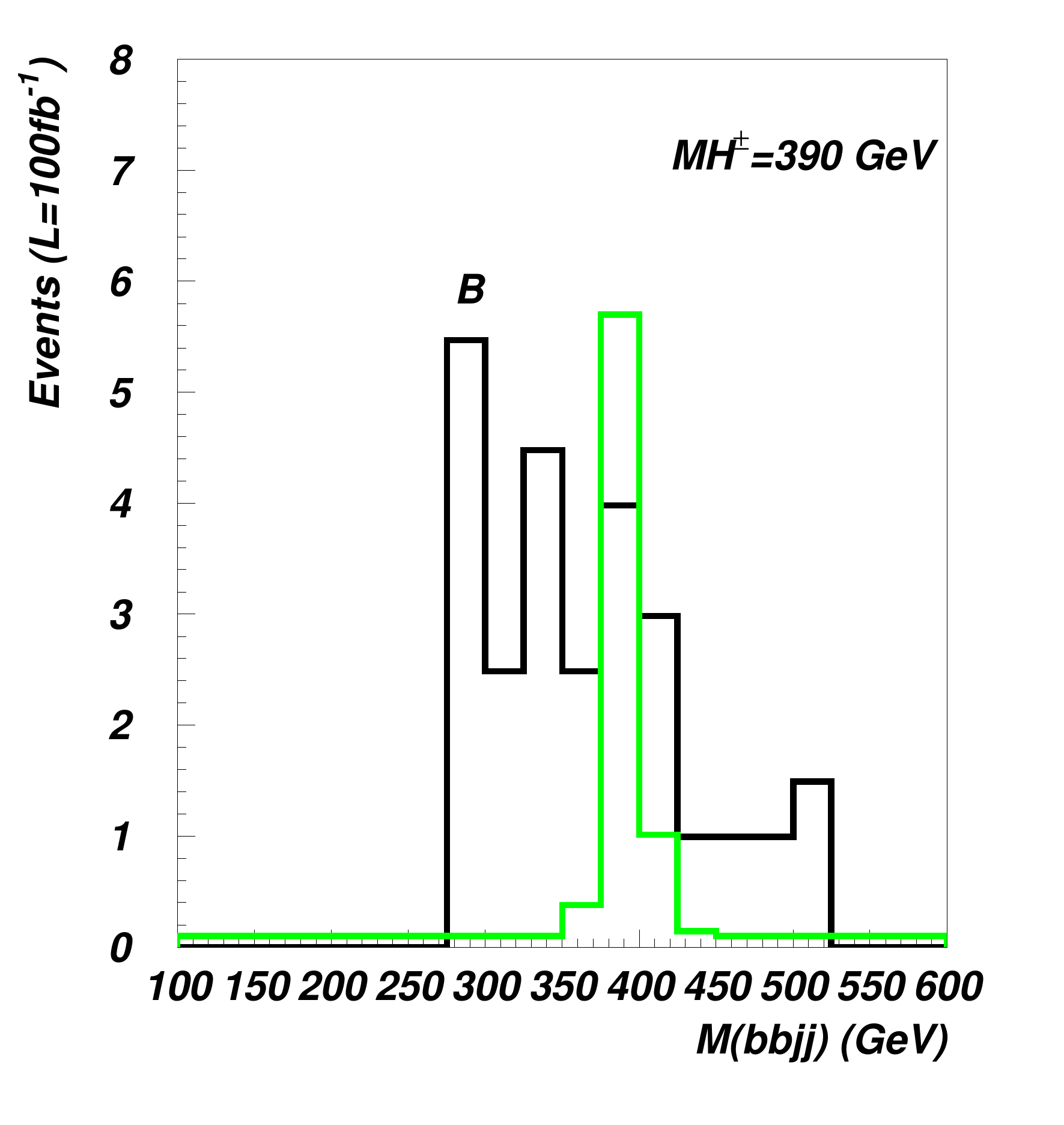}
\caption{Point $P_7$. Similar to fig.~\ref{Pi_2}. \label{Pi_7}}
\end{figure}

The invariant mass distributions for the points $P_2$, $P_3$, $P_4$, $P_5$, and $P_7$ are plotted in Figs.~\ref{Pi_2}--\ref{Pi_7}, each for two values of the charged Higgs mass. Table~\ref{summary_cuts} collects the results for the $S/\sqrt{B}$ analysis for all points.

\begin{table}[ht]
\begin{center}
\scalebox{0.8}{
\begin{tabular}{|c||c|c||c|c|}
\hline
 &
\multicolumn{2}{|c||}{$M_{H^\pm}=310$ GeV} & \multicolumn{2}{|c||}{$M_{H^\pm}=390$ GeV} \\ \hline
 & Events & $S/ \sqrt{B}$ & Events & $S/\sqrt{B}$ \\ 

\hline
$t\overline{t}$ & \multicolumn{4}{|c|}{24.9} \\ \cline{2-5}
peak & $11.9$ & $-$ & $9.9$ & $-$ \\
\hline
$P_1$ & $3.8$ & $0.8$ & $-$ & $-$   \\ 
   peak & $2.6$ & $0.8$ & $-$ & $-$ \\
\hline
$P_2$ & $4.7$ & $1.0$ & $8.8$ & $1.8$ \\ 
   peak & $3.3$ & $1.0$ & $7.3$ & $2.3$ \\
\hline
$P_3$ & $11.3$ & $2.3$ & $22.0$ & $4.4$ \\ 
   peak & $7.7$ & $2.3$ & $17.2$ & $5.4$ \\
\hline
$P_4$ & $10.0$ & $2.0$ & $20.3$ & $4.1$ \\ 
   peak & $7.8$ & $2.3$ & $16.0$ & $5.1$ \\
\hline
$P_5$ & $21.1$ & $4.2$ & $30.2$ & $6.1$ \\ 
   peak & $13.9$ & $4.1$ & $25.0$ & $7.9$  \\ 
\hline
$P_6$ & $14.0$ & $2.8$ & $-$ & $-$ \\ 
   peak & $9.4$ & $2.8$ & $-$ & $-$ \\ 
\hline
$P_7$ & $3.1$ & $0.6$ & $7.4$ & $1.5$ \\ 
   peak & $2.8$ & $0.8$ & $7.3$ & $2.3$ \\ 
\hline
$P_8$ & $1.2$ & $0.2$ & $-$ & $-$ \\ 
   peak & $1.2$ & $0.4$ & $-$ & $-$ \\ 
\hline
\end{tabular}}
\end{center}
\caption{Surviving events and their significance after the single cut of eq.~(\ref{cut_sg}) and after the peak selection of eq.~(\ref{cut_peak}), for all points of table~\ref{points}, except $P_9$ and $P_{10}$. \label{summary_cuts}}
\end{table}

As regards $P_1$, $P_6$ and $P_8$, no choice of allowed $M_{H^\pm}$ produces any appreciable signal after the whole set of cuts, hence we will not discuss them any further.

From fig.~\ref{Pi_2} we learn that a choice of $\tan\beta=1$ (though a rather low value of $M_{2}$ disfavours the production cross section) is enough to produce a visible signal, even for $M_{H^\pm}=310$ GeV with the signal peak lying over the background.
However, it is clear that the signal suffers from the selection cuts, and this critical situation is eased up only when $M_{H^\pm}=390$ GeV. In fact,
for higher allowed $M_{H^\pm}$ masses, the signal
contains a conspicuous number of events ($\sim 10$ in the peak bin), and it is always above the background.

If we consider the points $P_3$, $P_4$ and $P_5$ the same circumstances occur: the higher $M_{H^\pm}$, the more visible and clear the signal with respect to the background. Moreover, from $P_3$ ($M_2=350$ GeV) to $P_5$ ($M_2=450$ GeV) the production cross section increases, as shown in figs.~\ref{Pi_3}-\ref{Pi_5}.

If we consider the point $P_7$ (which has $\tan\beta=2$) in fig.~\ref{Pi_7}, we do not note any change from the previous considerations: even when $\tan\beta$ grows we still have a considerable production cross section in the allowed parameter space, and the signal would be observable with respect to the background, at least for $M_{H^\pm}=390$ GeV.

\section{Possible future scenarios}
\label{Sec:Scenarios}
\setcounter{equation}{0}
We shall here discuss possible future experimental developments,
and consider their implications for the model, in particular
for the proposed benchmarks.
The basic question is of course: which experimental efforts are
required to exclude the 2HDM altogether? We shall below adress a
couple of LHC-related aspects of this question.

\subsection{Higher and more constrained rates for $gg\to H_1\to\gamma\gamma$}
Several authors have recently argued that the LHC experiments
point to an overall rate for $pp\to H\to\gamma\gamma$ that is somewhat high
compared to the SM prediction. In our parameter scans in
section~\ref{Sec:bounds} we generously allowed the ratio
$R_{\gamma\gamma}$ of Eq.~(\ref{Eq:R_gammagamma}) to satisfy
$0.5\leq R_{\gamma\gamma}\leq2.0$. We shall here briefly comment on how the parameter
space is further constrained when we only allow the upper range:
\begin{equation} \label{Eq:R_gammagamma-tight}
1.5\leq R_{\gamma\gamma}\leq2.0
\end{equation}

\begin{figure}[htb] 
 \includegraphics[angle=0,width=0.9\textwidth
  ]{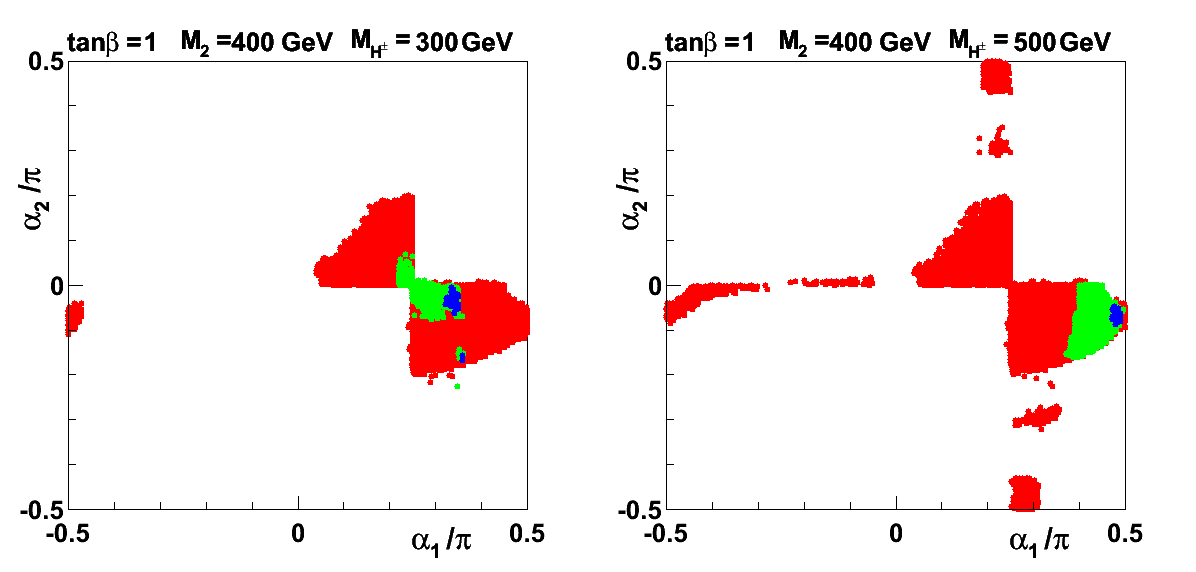}
\caption{Superimposed on Fig.~\ref{Fig:alpha12-1-400}, we show (in blue)
  the remaining allowed regions in the $\alpha_1$--$\alpha_2$
  parameter space, imposing higher rates for the $\gamma\gamma$
  channel, Eq.~(\ref{Eq:R_gammagamma-tight}), for
  $M_2=400~\text{GeV}$ and the additional parameters
  given in Eq.~(\ref{Eq:plot-params}).
\label{Fig:alpha12-tight}}
\end{figure}

For the case shown in Fig.~\ref{Fig:alpha12-1-400}, namely
$\tan\beta=1$ and $M_2=400~\text{GeV}$ (and two values of $M_{H^\pm}$), we show in
Fig.~\ref{Fig:alpha12-tight} how the LHC-allowed region gets
constrained. 
Such a development can have dramatic consequences for the
model: benchmark points $P_1$, $P_6$, $P_7$ and $P_8$ would be
excluded.

\begin{figure}[!htb] 
\includegraphics[angle=0,width=0.7\textwidth
  ]{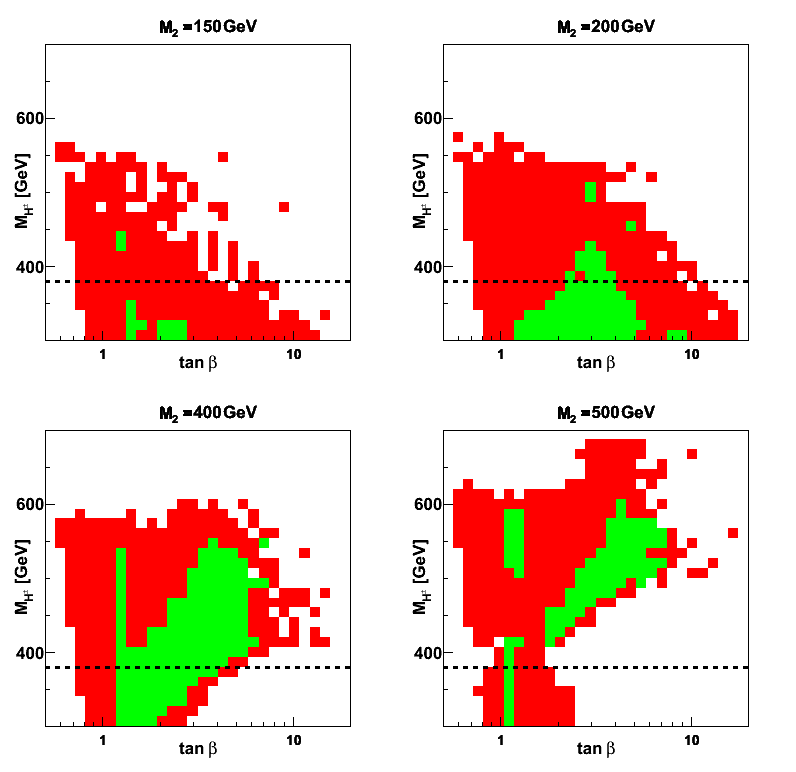}
 \caption{Allowed regions in the $\tan\beta$--$M_{H^\pm}$
  parameter space, without (red) and with (green) the LHC constraints,
  for the range in $R_{\gamma\gamma}$ given by Eq.~(\ref{Eq:R_gammagamma-tight}),
  and four values of $M_2$,
  as indicated.
The dashed lines show the recent bound at 380~GeV \cite{Hermann:2012fc}.
\label{Fig:tanbeta-mh_ch-tight}}
\end{figure}

In figure~\ref{Fig:tanbeta-mh_ch-tight} we show the remaining allowed
regions in the $\tan\beta$--$M_{H^\pm}$ plane, which exhibits a
significant reduction, as compared with
figure~\ref{Fig:tanbeta-mh_ch}. While the scans have limited
statistics, there is an indication that the remaining allowed
parameter space starts fragmenting into disconnected regions.

\subsection{Tightened upper bound on $gg\to H_{2,3}\to W^+W^-$ (and  $ZZ$)}

\begin{figure}[htb] 
 \includegraphics[angle=0,width=0.9\textwidth
  ]{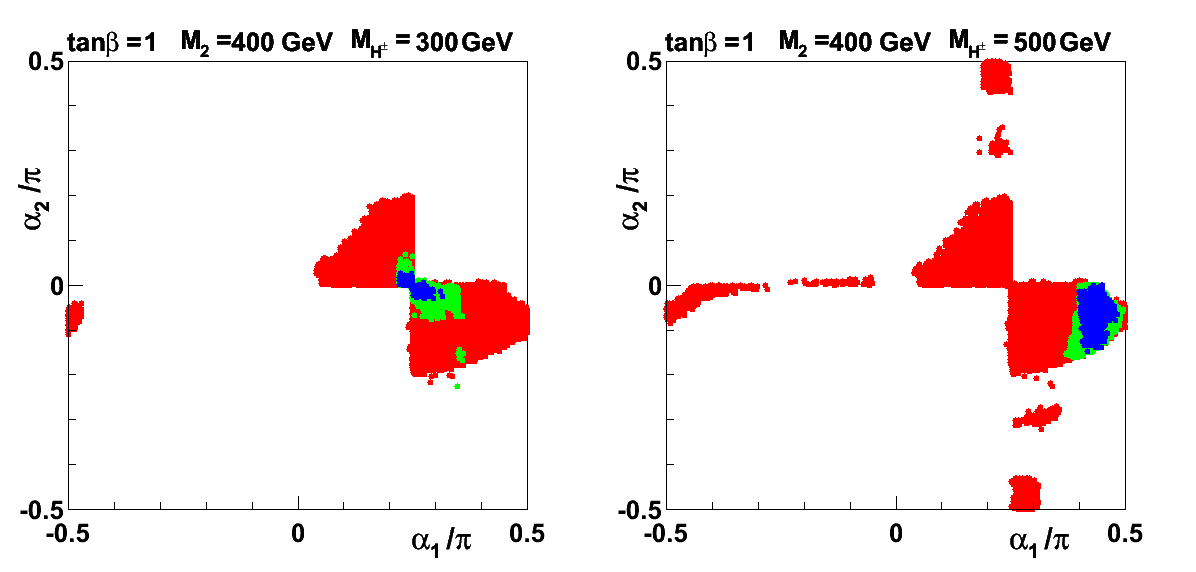}
\caption{Superimposed on Fig.~\ref{Fig:alpha12-1-400}, we show (in blue)
  the remaining allowed regions in the $\alpha_1$--$\alpha_2$
  parameter space, assuming that the upper bound on an SM-like Higgs
  is tightened by a factor 0.5 for masses in the range 130--600~GeV, for
  $M_2=400~\text{GeV}$ and the additional parameters
  given in Eq.~(\ref{Eq:plot-params}).
\label{Fig:alpha12-reducedWW}}
\end{figure}

Presumably, the search for an SM-like Higgs will continue in the mass
range from around 130~GeV and up. To a first (rough) approximation, a Higgs in
this mass region is produced via gluon fusion, and decays via $WW$ (or $ZZ$)
bosons. Assuming these upper bounds are tightened, it is interesting
to see how the allowed parameter space behaves. In
Fig.~\ref{Fig:alpha12-reducedWW} we show how the allowed parameter
space shrinks if we assume that the upper bound on a Higgs-like
particle, represented by the quantity $R_{ZZ}$ of Eq.~(\ref{Eq:R_ZZ}), is
lowered by a factor 0.5.
Again, we see a rather dramatic impact of such a development.

\section{Conclusions}
\label{Sec:Conclusions}
\setcounter{equation}{0}
For the channel $pp\to H^\pm W^\mp\to W^+W^-b\bar b$, we have
established a set of 7 benchmarks for the CP-violating 2HDM with
type-II Yukawa interactions. These points all have
$M_1=125~\text{GeV}$, low $\tan\beta$,
they all violate CP, and allow for a range of charged-Higgs masses.
A set of cuts is proposed, that will reduce the $t\bar t$ background
to a tolerable level, and allow for the detection of a signal in the
$WW\to jj\ell\nu$ channel. Some of the proposed benchmark points lead
to enhanced $H^\pm W^\mp$ production cross sections due to resonant
production via $H_2$ or $H_3$ in the $s$-channel.

Most of the proposed points are in the interior of some allowed domain in the
$\alpha$ space, and thus robust with respect to minor modifications of
the experimental constraints.
Some of the benchmark points are vulnerable to a higher value of
$R_{\gamma\gamma}$. However, the points $P_2$, $P_3$, $P_4$ and $P_5$ are not endangered.

It should also be noted that the proposed channel only benefits from favourable production cross sections and
branching ratio at low values of $\tan\beta$. In this region, the charged Higgs
mass is constrained to the range $\sim380$--$470~\text{GeV}$.

\vspace*{10mm} \noindent {\bf Acknowledgements.}  
GMP would like to thank Guido Macorini, Jae-Hyeon Park, Alexander Pukhov and Dominik St\"ockinger for helpful discussions. LB would like to thank Andrea Banfi for helpful discussion about cuts. \\
LB has been supported by the Deutsche Forschungsgemeinschaft through
the Research Training Group GRK\,1102 \textit{Physics of Hadron Accelerators}.
The work of AL, PO and MP has been supported by the Research Council of Norway.
SM and GMP acknowledge partial financial support through the NExT Institute.
The work of GMP has also been supported by the German Research Foundation DFG through Grant No.\ STO876/2-1 and by BMBF Grant No.\ 05H09ODE, the WUN Research Mobility Programme and the Research Council of Norway.

\appendix
\section{Appendix. The decoupling limit}
\setcounter{equation}{0}
\renewcommand{\thesection}{A}
\label{appe:a}

We shall here explore the so-called decoupling limit, which has been
studied for the CP-conserving case in \cite{Gunion:2002zf}, where 
\begin{equation}
M_{H^\pm}\sim M_3\sim M_2\gg M_1.
\end{equation}
We shall see that the large masses will all be of order $\mu$.
Note, however, that this discussion disregards the constraints of
positivity, unitarity, etc., that are discussed in section~\ref{Sec:bounds}.

\subsection{$\tan\beta$ of ${\cal O}(1)$}

For definiteness, we substitute
\begin{equation}
M_3=M_2=M
\end{equation}
into the expressions for the $\lambda$'s \cite{ElKaffas:2007rq},
require them all to be small, and also neglect terms of order $M_1^2$
compared to $M^2$. The conditions related to the different $\lambda$'s are:
\begin{subequations} \label{Eq:decoupling}
\begin{alignat}{2}
&\lambda_1: &\quad \label{Eq:decoupling-lambda1}
(c_1^2s_2^2+s_1^2)M^2&\simeq s_\beta^2\mu^2, \\
&\lambda_2: &\quad \label{Eq:decoupling-lambda2}
 (s_1^2s_2^2+c_1^2)M^2&\simeq c_\beta^2\mu^2, \\
&\lambda_3: &\quad
\frac{c_1s_1}{c_\beta s_\beta}c_2^2M^2&\simeq 2M_{H^\pm}^2-\mu^2, \\
&\lambda_4: &\quad \label{Eq:decoupling-lambda4}
c_2^2M^2&\simeq 2M_{H^\pm}^2-\mu^2, \\
&\Re\lambda_5: &\quad \label{Eq:decoupling-lambda5re}
c_2^2M^2&\simeq \mu^2, \\
&\Im\lambda_5: &\quad \label{Eq:decoupling-lambda5im}
(c_\beta c_1+s_\beta s_1) c_2s_2&\simeq 0.
\end{alignat}
\end{subequations}

Adding Eqs.~(\ref{Eq:decoupling-lambda1}), (\ref{Eq:decoupling-lambda2}) and (\ref{Eq:decoupling-lambda5re}), we get
\begin{equation}
M^2\simeq\mu^2,
\end{equation}
as anticipated.
Substituting back into Eqs.~(\ref{Eq:decoupling-lambda5re}) and (\ref{Eq:decoupling-lambda4}), we find
\begin{equation}
c_2\simeq1,
\end{equation}
and
\begin{equation}
M_{H^\pm}^2\simeq\mu^2\simeq M^2.
\end{equation}

The last equation, Eq.~(\ref{Eq:decoupling-lambda5im}), provides two
solutions, either
\begin{equation}
\cos(\beta-\alpha_1)\simeq0,\quad\text{or }
\sin(2\alpha_2)\simeq0.
\end{equation} 
We note that the angle $\alpha_3$ does not enter in these asymptotic
conditions (\ref{Eq:decoupling}), and that they are all satisfied for
\begin{equation} \label{Eq:decouple-summary}
\alpha_1\sim\beta, \quad \alpha_2\sim0, \quad \alpha_3\text{ arbitrary.}
\end{equation}

\subsection{Large $\tan\beta$}
The case of large values of $\tan\beta$ requires special
attention. Because of over-all factors which were left out in
Eq.~(\ref{Eq:decoupling}), the first on them,
Eq.~(\ref{Eq:decoupling-lambda1}) must be satisfied to a much higher
degree than the others (a factor $1/c_\beta^2$ is involved). This means that the expression
\begin{equation}
c_1^2s_2^2+s_1^2
\end{equation}
must be maximised. This requires 
\begin{equation}
\sin\alpha_1=\pm1, \quad
\sin\alpha_2=0,
\end{equation}
consistent with Eq.~(\ref{Eq:decouple-summary}) (and with $H_2$ or
$H_3$ being odd under CP), or
\begin{equation}
\sin\alpha_1=0, \quad
\sin\alpha_2=\pm1.
\end{equation}
The latter solution (which corresponds to $H_1$ being odd under CP) is not contained in Eq.~(\ref{Eq:decouple-summary}).

\bibliographystyle{h-physrev5}
\bibliography{biblio}

\end{document}